\def\longbuild{}

\documentclass[conference]{IEEEtran}

\usepackage[T1]{fontenc}

\usepackage{graphicx}
\usepackage{array}
\usepackage{booktabs}
\usepackage{tabularx}
\newcolumntype{Y}[1]{>{\hsize=#1\hsize\raggedright\arraybackslash}X}
\usepackage{siunitx}
\AtBeginDocument{\DeclareSIPrefix{\micro}{\text{\textmu}}{-6}}
\usepackage{amsmath}
\usepackage{mathtools}
\usepackage[newfloat=true]{minted}
\usepackage{caption}
\usepackage{subcaption}
\usepackage[listings, xparse, minted, skins, breakable]{tcolorbox}
\usepackage{hyperref}
\hypersetup{hidelinks}
\usepackage{bookmark}
\usepackage[capitalize]{cleveref}
\usepackage{cite}
\newif\iflong
\ifdefined\longbuild\longtrue\fi
\newcommand{\longonly}[1]{\iflong#1\fi}
\newcommand{\shortonly}[1]{\iflong\else#1\fi}
\usepackage{svg}
\usepackage{pgfplots}
\pgfplotsset{compat=1.18}
\usetikzlibrary{patterns}
\usepgfplotslibrary{external}
\DeclareSIUnit{\sample}{S}

\makeatletter

\pgfkeys{
  /scopeplot/.cd,
  title/.store in=\scp@title,       title={},
  file/.store in=\scp@file,         file={},
  file2/.store in=\scp@filetwo,     file2={},      % optional second trace
  skip/.store in=\scp@skip,         skip=0,        % header lines to discard
  colsep/.store in=\scp@colsep,     colsep=comma,
  vscale/.store in=\scp@vscale,     vscale=1,      % V per division
  vcenter/.store in=\scp@vcenter,   vcenter=0,     % V at the window centre
  hscale/.store in=\scp@hscale,     hscale=1e-9,   % s per division
  hpos/.store in=\scp@hpos,         hpos=0,        % s at the trigger fraction
  trigfrac/.store in=\scp@trigfrac, trigfrac=0.5,
  hdiv/.store in=\scp@hdiv,         hdiv=10,
  vdiv/.store in=\scp@vdiv,         vdiv=10,
  minor/.store in=\scp@minor,       minor=5,       % subdivisions per division
  nth/.store in=\scp@nth,           nth=1,         % keep every nth sample
  width/.store in=\scp@width,       width=0.86\columnwidth,
  height/.store in=\scp@height,     height=115pt,
  input/.store in=\scp@input,       input={},      % coupling and termination
  bw/.store in=\scp@bw,             bw={},         % Hz
  rate/.store in=\scp@rate,         rate={},       % S/s, hardware rate
}

\newcommand{\scp@gridstyle}[2]{%
  \ifnum#1=0 solid\else
  \ifnum#1=#2 solid\else
  \ifnum\numexpr2*#1\relax=#2 densely dotted\else
  loosely dotted\fi\fi\fi}

\newcommand{\scp@eval}[3]{%
  \begingroup
  \pgfkeys{/pgf/fpu, /pgf/fpu/output format=#3}%
  \pgfmathparse{#2}%
  \expandafter\endgroup\expandafter\def\expandafter#1\expandafter{\pgfmathresult}}

\newcommand{\scp@qty}[2]{%
  \qty[exponent-mode=engineering, prefix-mode=combine-exponent,
       per-mode=symbol, round-mode=figures, round-precision=4,
       round-pad=false, drop-zero-decimal]{#1}{#2}}

\newcommand{\scopeplot}[1]{%
  \begingroup
  \begin{tikzpicture}
  \pgfkeys{/scopeplot/.cd,#1}%
  \scp@eval{\scp@ymin}{\scp@vcenter-0.5*\scp@vdiv*\scp@vscale}{sci}%
  \scp@eval{\scp@ymax}{\scp@vcenter+0.5*\scp@vdiv*\scp@vscale}{sci}%
  \scp@eval{\scp@tmin}{\scp@hpos-\scp@trigfrac*\scp@hdiv*\scp@hscale}{sci}%
  \scp@eval{\scp@tmax}{\scp@tmin+\scp@hdiv*\scp@hscale}{sci}%
  \scp@eval{\scp@gnd}{0.5-\scp@vcenter/(\scp@vdiv*\scp@vscale)}{fixed}%
  \pgfmathtruncatemacro{\scp@nxm}{\scp@hdiv*\scp@minor-1}%
  \pgfmathtruncatemacro{\scp@nym}{\scp@vdiv*\scp@minor-1}%
    \begin{axis}[
      width=\scp@width, height=\scp@height, scale only axis,
      title style={align=center, font=\footnotesize}, title={\scp@title},
      axis lines=none, clip=false,
      xmin=\scp@tmin, xmax=\scp@tmax, ymin=\scp@ymin, ymax=\scp@ymax,
      ]
      \foreach \i in {0,...,\scp@hdiv}{%
        \pgfmathsetmacro{\ff}{\i/\scp@hdiv}%
        \edef\scp@tmp{\noexpand\draw[\scp@gridstyle{\i}{\scp@hdiv}]
          (rel axis cs:\ff,0) -- (rel axis cs:\ff,1);}\scp@tmp}
      \foreach \j in {0,...,\scp@vdiv}{%
        \pgfmathsetmacro{\ff}{\j/\scp@vdiv}%
        \edef\scp@tmp{\noexpand\draw[\scp@gridstyle{\j}{\scp@vdiv}]
          (rel axis cs:0,\ff) -- (rel axis cs:1,\ff);}\scp@tmp}
      \foreach \k in {1,...,\scp@nxm}{%
        \pgfmathsetmacro{\ff}{\k/(\scp@hdiv*\scp@minor)}%
        \ifnum\numexpr\k-\scp@minor*(\k/\scp@minor)\relax=0
          \def\ch{2}\def\cw{4}\def\bl{3.5}\else
          \def\ch{1}\def\cw{2}\def\bl{2}\fi
        \edef\scp@tmp{%
          \noexpand\draw[very thin] (rel axis cs:\ff,0.5)
            ++(0pt,\ch pt) -- ++(0pt,-\cw pt);
          \noexpand\draw[very thin] (rel axis cs:\ff,0) -- ++(0pt,\bl pt);
          \noexpand\draw[very thin] (rel axis cs:\ff,1) -- ++(0pt,-\bl pt);%
        }\scp@tmp}
      \foreach \k in {1,...,\scp@nym}{%
        \pgfmathsetmacro{\ff}{\k/(\scp@vdiv*\scp@minor)}%
        \ifnum\numexpr\k-\scp@minor*(\k/\scp@minor)\relax=0
          \def\ch{2}\def\cw{4}\def\bl{3.5}\else
          \def\ch{1}\def\cw{2}\def\bl{2}\fi
        \edef\scp@tmp{%
          \noexpand\draw[very thin] (rel axis cs:0.5,\ff)
            ++(\ch pt,0pt) -- ++(-\cw pt,0pt);
          \noexpand\draw[very thin] (rel axis cs:0,\ff) -- ++(\bl pt,0pt);
          \noexpand\draw[very thin] (rel axis cs:1,\ff) -- ++(-\bl pt,0pt);%
        }\scp@tmp}
      \ifx\scp@filetwo\@empty
        \addplot[mark=none, opacity=0.9, each nth point=\scp@nth,
          filter discard warning=false, unbounded coords=discard,
          restrict x to domain=\scp@tmin:\scp@tmax,
          restrict y to domain*=\scp@ymin:\scp@ymax]
          table[col sep=\scp@colsep, x index=0, y index=1,
                skip first n=\scp@skip] {\scp@file};
      \else
        \addplot[mark=none, densely dotted, each nth point=\scp@nth,
          filter discard warning=false, unbounded coords=discard,
          restrict x to domain=\scp@tmin:\scp@tmax,
          restrict y to domain*=\scp@ymin:\scp@ymax]
          table[col sep=\scp@colsep, x index=0, y index=1,
                skip first n=\scp@skip] {\scp@file};
        \addplot[mark=none, each nth point=\scp@nth,
          filter discard warning=false, unbounded coords=discard,
          restrict x to domain=\scp@tmin:\scp@tmax,
          restrict y to domain*=\scp@ymin:\scp@ymax]
          table[col sep=\scp@colsep, x index=0, y index=1,
                skip first n=\scp@skip] {\scp@filetwo};
      \fi
      \ifdim\scp@gnd pt>0pt \ifdim\scp@gnd pt<1pt
        \draw (rel axis cs:0,\scp@gnd) -- ++(-5pt,0pt) -- ++(0pt,-3pt);
        \draw (rel axis cs:0,\scp@gnd) ++(-7.5pt,-3pt)
          -- ++(5pt,0pt) -- ++(-2.5pt,-5pt) -- cycle;
      \fi\fi
      \node at (rel axis cs:0,0) [anchor=north west, align=left]
        {\footnotesize\scp@qty{\scp@vscale}{\volt}/div};
      \node at (rel axis cs:0.5,0) [anchor=north, align=center]
        {\footnotesize\scp@input
         \ifx\scp@bw\@empty\else\\\footnotesize BW \scp@qty{\scp@bw}{\hertz}\fi};
      \node at (rel axis cs:1,0) [anchor=north east, align=right]
        {\footnotesize\scp@qty{\scp@hscale}{\second}/div
         \ifx\scp@rate\@empty\else
           \\\footnotesize\scp@qty{\scp@rate}{\sample\per\second}\fi};
    \end{axis}
  \end{tikzpicture}%
  \endgroup
}

\makeatother

\setminted{fontsize=\footnotesize}
\setminted{linenos=true}
\setminted{numbersep=4pt}
\setminted{breaklines=true}
\setminted{mathescape=true}
\definecolor{codebg}{rgb}{0.95,0.95,0.95}

\SetupFloatingEnvironment{listing}{name=Listing, listname=List of Listings}
\crefname{listing}{listing}{listings}
\Crefname{listing}{Listing}{Listings}

\tcbset{
  codecaption/.code={\gdef\codecaptiontext{#1}},
  codelabel/.code={\gdef\codelabeltext{#1}},
}
\newcommand{\codeaftercaption}{%
  \ifx\codecaptiontext\empty\else
    \par\nobreak\vskip\abovecaptionskip
    \captionof{listing}{\codecaptiontext}%
    \ifx\codelabeltext\empty\else\label{\codelabeltext}\fi
  \fi
  \par\vskip\baselineskip
}
\newtcblisting{code}[2][]{
  colback=codebg, colframe=codebg, left=0mm, right=0mm, top=0mm,
  bottom=0mm, leftrule=0mm, rightrule=0mm, toprule=0mm,
  bottomrule=0mm,
  breakable,listing only,
  before skip=\baselineskip,
  after skip=\baselineskip,
  break at=-\baselineskip/0pt,
  minted language={#2},
  nofloat,
  minipage boxed title,
  enlargepage flexible=2\baselineskip,
  codecaption={}, codelabel={},
  #1,
  after={\codeaftercaption},
}

\DeclarePairedDelimiter{\abs}{\lvert}{\rvert}

\begin{document}
% Abbreviate author lists in the bibliography (IEEEtranBSTCTL entry in biblio.bib).
\bstctlcite{IEEEexample:BSTcontrol}

\title{Error-Bounded Fixed-Point Design of Super-Sample-Rate IIR Filters for Real-Time Superconducting Qubit Flux Predistortion}

\author{

\IEEEauthorblockN{Matt Huszabianlou, Angelos Ioannou, Nirmalendu Bikash Patra, Anastasiia Butko, Gang Huang, and Irfan Siddiqi}

\IEEEauthorblockA{
Lawrence Berkeley National Laboratory\\
Berkeley, CA, USA\\
\{mhuszabianlou, aioannou, nbp, abutko, ghuang, isiddiqi\}@lbl.gov
}
}

\maketitle

\begin{abstract}
Flux-activated two-qubit gates in superconducting processors require correction of line distortion, which otherwise produces residual detuning, conditional-phase error, and leakage.
Prior demonstrations of cryoscope-based distortion calibration have generally used commercial solutions that specify the filter models but not their FPGA implementation.
These details matter for custom and open-source systems, where the fixed-point coefficient and datapath formats set how accurately the correction is realized, and the IIR feedback loop in its standard form does not meet timing at the required clock rate.
We derive the coefficient formats of the correction filters from a step response error tolerance, using bounds on the quantization-induced displacement of the poles and zeros over the physical parameter range of each distortion, and the accumulator format from the smallest input change that must remain resolvable at the output.
The same analysis gives the conditions under which quantization preserves filter stability.
We implement the resulting FIR/IIR cascade on the QubiC platform at a fabric clock rate of \SI{500}{MHz}, using a super-sample-rate structure with scattered look-ahead pipelining, which removes the feedback recursion from the critical path without changing the target transfer function.
A representative cascade for our \SI{1}{GS/s} flux lines, comprising an integrator, a second-order section, and a 20-tap FIR, consumes 88 DSP slices and adds \SI{162}{ns} of latency.
We confirm the hardware implementation by correcting characteristic bias-tee distortion, an intermediate step toward full cryoscope-based calibration.
\end{abstract}

\section{Introduction}

Transmon-based superconducting quantum processors frequently adopt an architecture in which the qubits, the couplers, or both are frequency-tunable by adjusting the magnetic flux through the transmon's SQUID\@.
Such tunability enables fast flux-activated two-qubit gates~\cite{rol2020cryoscope, glaser2024sensitivity}.
Non-parametric gates of this kind are driven by baseband pulses on the qubit flux lines, the coupler flux lines, or both, with either a rectangular envelope, usually with Gaussian-filtered edges to limit bandwidth, or a smooth one such as a Slepian or Fourier-series shape~\cite{martinis2014adiabatic, li2024doubletransmon, glaser2024sensitivity}.
Such lines typically incorporate a bias tee, which combines a low-noise DC bias with the fast pulse generated by the control system~\cite{hellings2025calibrating}.
The bias tee, together with various imperfections in the DAC and the flux line, distorts the waveform, so that what arrives at the transmon is no longer the intended pulse.
This distortion degrades gate fidelity: short-timescale distortion produces leakage out of the computational subspace, and long-timescale distortion makes the phase accumulated during a gate depend on the preceding pulse history~\cite{rol2020cryoscope}.
Numerous papers discuss calibration procedures for the filters used on qubit~\cite{rol2020cryoscope, jerger2019vna, guo2024universal, hellings2025calibrating, venkateswaran2026dpd} and coupler~\cite{li2025couplerpulse, zhang2025aswap} flux lines, and others mitigate distortion by shaping the pulse itself~\cite{glaser2024sensitivity, aggarwal2025transients}, but there is minimal discussion of the filter hardware implementation.
Most such work uses commercial arbitrary waveform generators, such as the Zurich Instruments HDAWG~\cite{zihdawg}, whose built-in real-time IIR and FIR predistortion filters are configured through model parameters, so the implementation is never exposed to the user.
For custom and open-source systems such as our QubiC project~\cite{xu2023qubic2, qubicweb}, however, these details are important, since implementing IIR filters with properly designed fixed-point quantization, saturation, and rounding behavior at high clock frequency and super-sample rate raises two distinct engineering challenges.
The coefficient and accumulator fixed-point formats determine how accurately the intended poles and zeros are realized, and translating a required correction accuracy into those formats is what allows a design to be sized for a given application.
Separately, the feedback recursion, evaluated directly, requires multiple interdependent multiply-add-round-saturate operations within a single clock cycle, each of which takes several cycles.

This paper makes the following technical contributions:

\begin{itemize}
  \item A fixed-point analysis that derives the coefficient formats from a step response error tolerance through bounds on the quantization-induced displacement of the poles and zeros over the physical parameter ranges, derives the accumulator format from the smallest input change that must remain resolvable at the output, and gives the conditions under which quantization preserves stability.
  \item A super-sample-rate FPGA realization of Parhi and Messerschmitt's pipelined block IIR~\cite{parhi1989pipeline, parhi1989block}, in which the transformed feedforward section reuses the polyphase FIR, each feedback loop includes rounding and saturation, and the transformed coefficients of every correction filter are given in closed form.
  \item A mapping onto the Xilinx DSP48E2 that keeps the datapath inside DSP columns and their dedicated cascade routing, performs convergent rounding and guaranteed saturation, and meets timing at \SI{500}{MHz}; each module in isolation has an Fmax of \SI{775}{MHz}, the maximum supported by the target device's DSP slices.
  \item A public release of the implementation upon publication.
\end{itemize}

\section{Related Work}\label{sec:related}

\subsection{Flux Distortion Calibration}

Prior work has concentrated on identifying the transfer function of the flux line and fitting a predistortion filter to its inverse.
Most of it rests on the cryoscope~\cite{rol2020cryoscope} technique, which uses the qubit as an in situ probe to measure the step response of its own flux line at the temporal resolution of the DAC\@.
Jerger et al.~\cite{jerger2019vna} give an independent alternative, treating the qubit as a vector network analyzer to measure the transfer function in the frequency domain.
Hellings et al.~\cite{hellings2025calibrating} pair cryoscope with time-resolved qubit spectroscopy, which reaches \SI{100}{\micro\second} at coarser time resolution, to cover distortion from nanoseconds to tens of microseconds.
Guo et al.~\cite{guo2024universal} model the line by its system function rather than its step or frequency response, extracting the poles and zeros by an optimization initialized from an adjacent line of the same construction.
Tunable couplers usually have no dedicated readout, so their step response must be recovered indirectly, either through the coupler's strong coupling to neighboring qubits~\cite{li2025couplerpulse} or through an adiabatic swap between coupler and qubit~\cite{zhang2025aswap}.
In each case the filter is specified as a model, with its realization left to the control hardware.

\subsection{High-Throughput IIR Architectures}

Two families of transformations raise the throughput of a recursive filter beyond the limit set by its feedback loop.
Look-ahead pipelining rewrites the recursion so that each output depends only on outputs at least \(J\) samples in the past, augmenting the transfer function with canceling pole-zero pairs.
Clustered look-ahead~\cite{loomis1984highspeed} keeps the transformed feedback taps contiguous but does not guarantee that the added poles lie inside the unit circle~\cite[Sec.~IV-A]{parhi1989pipeline}; Lim and Liu~\cite{lim1992pipelined} find the minimum augmentation that preserves stability.
Scattered look-ahead~\cite{parhi1989pipeline} constrains the transformed denominator to powers of \(z^{-J}\), which places the added poles at the radii of the original poles, so stability is preserved for any \(J\).
Block processing produces \(M\) outputs per iteration~\cite{burrus1971block}, and Parhi and Messerschmitt~\cite{parhi1989block} combine \(M\)-parallel block processing with \(L\)-stage loop pipelining to multiply the sample rate by \(L M\).
Quantized coefficients cancel the added poles only inexactly, and Parhi~\cite{parhi1991finite} bounds the wordlength increase needed to keep the output error of the transformed filter at that of the original.
Scattered look-ahead has also been used in the FPGA front end of a physics instrument, the radio-interference notch filters of the Auger Engineering Radio Array, where the coefficients are converted to fixed point without a stated wordlength or accuracy criterion~\cite{szadkowski2015aera}.
Neither work derives wordlengths from an application-level accuracy requirement.

\section{Distortion Models and Corrections}\label{sec:distortion}

\begin{figure}\centering
  \includesvg[pretex=\footnotesize]{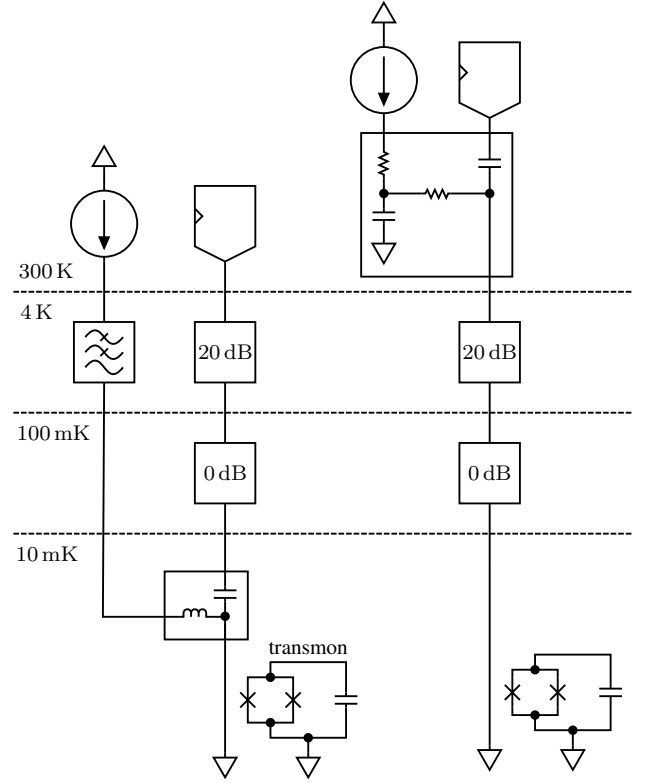}
  \caption{Typical configurations of qubit and coupler flux tuning lines. Two common variants include (left) an inductive bias tee in the mixing chamber~\cite{guo2024universal} and (right) a resistive bias tee at room temperature~\cite{hellings2025calibrating}.}\label{fig:line}
\end{figure}

We assume the flux tuning line, \cref{fig:line}, can be modeled as a linear time-invariant (LTI) system.
Its transfer function can then be measured, and the inverse of that transfer function fit to a cascade of digital filters that cancels the distortion.
In this section, \(G(z)\) denotes the transfer function of the distortion, \(H(z) \coloneq G(z)^{-1}\) denotes the transfer function of the associated compensation, and \(q_{i}\) and \(p_{i}\) denote the zeros and poles, respectively, of \(H(z)\).

\subsection{High-Pass Droop}\label{sec:high_pass_droop}

The dominant form of line distortion is the droop caused by the blocking capacitor in the bias tee together with the source and load resistances~\cite{rol2020cryoscope, hellings2025calibrating}.
An RC high-pass filter with time constant \(\tau = RC\), where \(R\) is the total series resistance (source plus load) and \(C\) is the blocking capacitance, has the continuous-time transfer function
\begin{equation}\label{eq:highpass_s}
  G(s) = \frac{s \tau}{1 + s \tau},
\end{equation}
where \(s\) is the complex frequency variable.
On a DC-coupled line this term is absent and the integrator stage of \cref{fig:cascade} is bypassed.

We convert \cref{eq:highpass_s} to discrete time by the step-invariant transformation~\cite[Sec.~11.2.4]{oppenheim1997}.
The step response of \cref{eq:highpass_s} is \(g(t) = e^{-t/\tau}\), so with
\begin{equation}\label{eq:integrator_coeff}
  \rho \coloneq e^{-T_{s}/\tau},
\end{equation}
where \(T_{s}\) is the sample period, the sampled step response is \(g[n] = \rho^{n}\), whose \(z\)-transform is \(1/(1 - \rho z^{-1})\).
Throughout, \(\rho\) denotes the per-sample decay factor of a time constant \(\tau\).
Dividing by the \(z\)-transform of the discrete-time unit step, \(1/(1 - z^{-1})\), gives \(G(z) = (1 - z^{-1}) / (1 - \rho z^{-1})\), whose inverse, the predistortion correction, is
\begin{equation}\label{eq:integrator_transfer_function}
  H(z) = \frac{1 - \rho z^{-1}}{1 - z^{-1}}.
\end{equation}
The correction has a single zero \(q_{1} = \rho\) and a single pole \(p_{1} = 1\).

\Cref{eq:integrator_transfer_function} has unity gain well above the corner frequency, so no rescaling of the waveform is needed, but it has a pole at \(z = 1\) and thus infinite DC gain.
Therefore, one must take care to avoid saturating the integrator by employing net-zero pulses~\cite{rol2019netzero}, \cite[App.~F]{hellings2025calibrating}, or limiting the integral of unipolar DC inputs.
Saturation ensures the output does not wrap for nonconforming inputs, and convergent rounding keeps net-zero pulses from accumulating a bias.

% Derivation of \cref{eq:integrator_transfer_function}.
% The discrete-time unit step has \(X(z) = \sum_{n=0}^{\infty} z^{-n} = 1 / (1 - z^{-1})\).
% The continuous-time unit step has Laplace transform \(1/s\), so \(Y(s) = G(s)/s = \tau / (1 + s \tau) = 1 / (s + 1/\tau)\), whose inverse transform is \(g(t) = e^{-t/\tau}\).
% Sampling at \(t = n T_{s}\) gives \(g[n] = \rho^{n}\), with \(Z\)-transform \(\sum_{n=0}^{\infty} \rho^{n} z^{-n} = 1 / (1 - \rho z^{-1})\).
% \(G(z) = Y(z) / X(z) = (1 - z^{-1}) / (1 - \rho z^{-1})\), and \cref{eq:integrator_transfer_function} follows by inversion.
% The difference equation of \cref{eq:integrator_transfer_function} is \(y[n] = y[n-1] + x[n] - \rho\, x[n-1]\).

\subsection{Exponential Tail (Overshoot and Undershoot)}\label{sec:exponential_tail}

Attenuators, low-pass filters, lossy cable, and parasitic reactances contribute exponential settling tails~\cite[Sec.~III]{hellings2025calibrating}, \cite[Sec.~4]{foxen2019fluxsampling}.
A stable, overdamped circuit with \(N_{\tau}\) linearly independent reactive elements has the step response
\begin{equation}\label{eq:sum_exp}
  g(t) = \left(\alpha_{0} + \sum_{i=1}^{N_{\tau}} \alpha_{i} e^{-t/\tau_{i}}\right) u(t),
\end{equation}
where \(u(t)\) is the unit step, \(\tau_{i}\) are the time constants, \(\alpha_{i}\) are real weights, and \(\alpha_{0} = g(\infty)\) is the DC gain~\cite[App.~A]{hellings2025calibrating}.
The high-pass droop of \cref{sec:high_pass_droop} is the special case \(\alpha_{0} = 0\), \(N_{\tau}=1\).

Normalizing a single tail to unity DC gain gives \(g(t) = (1 + \alpha e^{-t/\tau}) u(t)\), where \(\alpha \coloneq \alpha_{1} / \alpha_{0} > -1\) is the amplitude coefficient; \(\alpha > 0\) is an overshoot and \(\alpha < 0\) an undershoot.
Applying the step-invariant transformation of \cref{sec:high_pass_droop} to \(g[n] = 1 + \alpha \rho^{n}\), with \(\rho\) from \cref{eq:integrator_coeff}, gives \(G(z) = [(1 + \alpha) - (\rho + \alpha) z^{-1}] / (1 - \rho z^{-1})\), whose inverse is
\begin{equation}\label{eq:tail_pre}
  H(z) = \kappa\,
                        \frac{1 - q_{1} z^{-1}}{1 - p_{1} z^{-1}},
\end{equation}
where
\begin{equation}\label{eq:tail_zeros_poles}
  \kappa \coloneq \frac{1}{1 + \alpha},
  \quad q_{1} \coloneq \rho,
  \quad p_{1} \coloneq \frac{\rho + \alpha}{1 + \alpha}.
\end{equation}
\(\kappa\) normalizes the DC gain of \cref{eq:tail_pre} to unity.

% The difference equation of \cref{eq:tail_pre} is \(y[n] = p_{1}\, y[n-1] + \kappa \left(x[n] - \rho\, x[n-1]\right)\).

\subsection{Damped Oscillation}\label{sec:damped_oscillation}

\Cref{eq:sum_exp} presumes the natural frequencies of the circuit are all real~\cite[App.~A]{hellings2025calibrating}, which fails for an underdamped section such as a bias tee whose inductance resonates with stray capacitance~\cite{guo2024universal}, or a pair of impedance mismatches whose standing wave appears as a broad resonance in the measured transfer function~\cite{jerger2019vna}, \cite[App.~G]{aggarwal2025transients}.
Such a section contributes a decaying oscillation \(2 \alpha_{r} e^{-t/\tau} \cos(2 \pi t / T_{\mathrm{osc}} + \phi)\) to the step response, with amplitude \(2 \alpha_{r}\), phase \(\phi\), envelope time constant \(\tau\), and period \(T_{\mathrm{osc}}\)~\cite[Eq.~4]{guo2024universal}, giving it a quality factor \(Q \coloneq \pi \tau / T_{\mathrm{osc}}\).
This appears in \(G(z)\) as a conjugate pole pair at \(z = \rho e^{\pm j \theta}\), with \(\rho\) from \cref{eq:integrator_coeff} and \(\theta = 2 \pi T_{s} / T_{\mathrm{osc}}\), contributing the quadratic factor \(1 - 2 \rho \cos\theta\, z^{-1} + \rho^{2} z^{-2}\) to its denominator.

The residues determine the numerator of \(G(z)\) once the partial-fraction terms are placed over their common denominator~\cite{guo2024universal}.
Applying the step-invariant transformation of \cref{sec:high_pass_droop} to \(g[n] = 1 + 2 \alpha_{r} \rho^{n} \cos(n \theta + \phi)\), normalized to unity DC gain as in \cref{sec:exponential_tail}, gives
\begin{equation}\label{eq:osc_z}
  G(z) = \frac{c_{0} + c_{1} z^{-1} + c_{2} z^{-2}}
              {1 - 2 \rho \cos\theta\, z^{-1} + \rho^{2} z^{-2}},
\end{equation}
with
\begin{subequations}\label{eq:osc_num}
\begin{align}
  c_{0} &= 1 + 2 \alpha_{r} \cos\phi,\\
  c_{1} &= -2 \left[\rho \cos\theta
           + \alpha_{r} \left(\cos\phi + \rho \cos(\theta - \phi)\right)\right],\\
  c_{2} &= \rho^{2} + 2 \alpha_{r} \rho \cos(\theta - \phi).
\end{align}
\end{subequations}
The correction is \(1/G(z)\), which exchanges the poles and zeros of \(G(z)\), giving the second-order section
\begin{equation}\label{eq:sos}
  H(z) = \kappa\,
    \frac{(1 - q_{1} z^{-1})(1 - q_{2} z^{-1})}
         {(1 - p_{1} z^{-1})(1 - p_{2} z^{-1})}.
\end{equation}
Here the zeros are the resonance pair, \(q_{1,2} = \rho e^{\pm j \theta}\), and the poles are the roots of \(c_{0} z^{2} + c_{1} z + c_{2}\), satisfying \(p_{1} + p_{2} = -c_{1}/c_{0}\) and \(p_{1} p_{2} = c_{2}/c_{0}\).
The gain \(\kappa = 1/c_{0}\) is what remains after factoring the leading coefficient out of the numerator of \cref{eq:osc_z}, since \(c_{0} + c_{1} z^{-1} + c_{2} z^{-2} = c_{0} (1 - p_{1} z^{-1})(1 - p_{2} z^{-1})\).

\subsection{Impedance Discontinuity Reflections (Bounce)}\label{sec:bounce}

An impedance discontinuity returns a delayed copy of the signal, \(G(z) = 1 + \alpha_{e} z^{-D}\), where \(\alpha_{e}\) is the relative echo amplitude and \(D\) is the echo delay in samples~\cite[App.~A]{hellings2025calibrating}.
The correction is
\begin{equation}\label{eq:bounce}
  H(z) = \frac{1}{1 + \alpha_{e} z^{-D}}
                      = \sum_{k=0}^{\infty} (-\alpha_{e})^{k} z^{-k D}.
\end{equation}

\subsection{Residual Short-Timescale Distortions}\label{sec:residual}

Finite DAC output bandwidth, cryogenic low-pass filters, the skin effect in semi-rigid coaxial cable, and impedance mismatches in the chip package and on-chip flux line leave residual distortion below a few tens of nanoseconds that has no compact pole-zero description~\cite{rol2020cryoscope, hellings2025calibrating}.
Additionally, we employ zero-order hold interpolation to resample the \SI{1}{GS/s} baseband signal to the DAC sample rate of \SI{8}{GS/s}, which produces image signals above the \SI{500}{MHz} Nyquist frequency and a sinc envelope with its first null at \SI{1}{GHz}.
An analog low-pass filter must be used to attenuate the image signals, but the FIR can perform the inverse sinc correction.
The correction is a non-parametric FIR, \(H(z) = \sum_{k=0}^{N_{b}-1} b_{k} z^{-k}\), whose \(N_{b}\) taps are fit numerically to the residual step response measured after the IIR stages~\cite{rol2020cryoscope, hellings2025calibrating, guo2024universal}.
The required tap count is the product of the correction span and the sample rate, so the \SIrange{20}{50}{ns} spans reported in these works correspond to 20 to 50 taps at our \SI{1}{GS/s} rate.
Because this residual is predominantly low-pass, its ideal inverse amplifies measurement noise toward the \SI{500}{MHz} Nyquist frequency, so the taps are fit to a bandwidth-limited target rather than an ideal impulse~\cite[App.~I]{hellings2025calibrating}, \cite{guo2024universal}.

\section{Filter Architecture}\label{sec:arch}

Our implementation is built from two configurable primitives: an FIR filter and an IIR filter.
Both parameterize the super-sample rate \(M\) and the feedforward tap count \(N_{b}\); the IIR additionally parameterizes the feedback tap count \(N_{a}\).
Both are fully pipelined, consuming and producing \(M\) samples per clock cycle, and both meet timing at QubiC's DSP clock frequency of \(f_{\mathrm{clk}} = \SI{500}{MHz}\) on a Xilinx UltraScale+ platform.
The resulting sample rate is \(f_{s} = M f_{\mathrm{clk}} = 1/T_{s}\).
All rounding is convergent (round half to even), and every datapath is guaranteed to saturate rather than overflow.
We specialize the IIR in three ways.
The integrator implements \(y[n] = y[n-1] + x[n] - \rho\, x[n-1]\), the difference equation of \cref{eq:integrator_transfer_function}, and is purpose-built to correct high-pass droop (\cref{sec:high_pass_droop}).
The first-order section (FOS) implements \(y[n] = b_{0} x[n] + b_{1} x[n-1] + a_{1} y[n-1]\) and handles overshoot and undershoot (\cref{sec:exponential_tail}).
The second-order section (SOS) implements \(y[n] = b_{0} x[n] + b_{1} x[n-1] + b_{2} x[n-2] + a_{1} y[n-1] + a_{2} y[n-2]\) and handles damped oscillations (\cref{sec:damped_oscillation}).

\begin{figure}\centering
  \includesvg{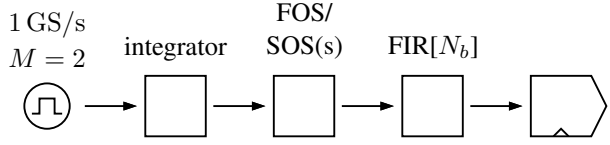}
  \caption{Filter cascade comprising an integrator, one or more first- or second-order sections, and an FIR of \(N_{b}\) taps.}\label{fig:cascade}
\end{figure}

We employ a filter cascade, as depicted in \cref{fig:cascade}, comprising an integrator, one or more first- or second-order sections, and an FIR of configurable length.
Our current configuration uses \(M=2\), for an overall sample rate of \(f_{s} = \SI{1}{GS/s}\).
This keeps rise times short while limiting FPGA DSP48E2 use, which grows quadratically with \(M\).

\subsection{FIR}

\begin{figure*}[!t]
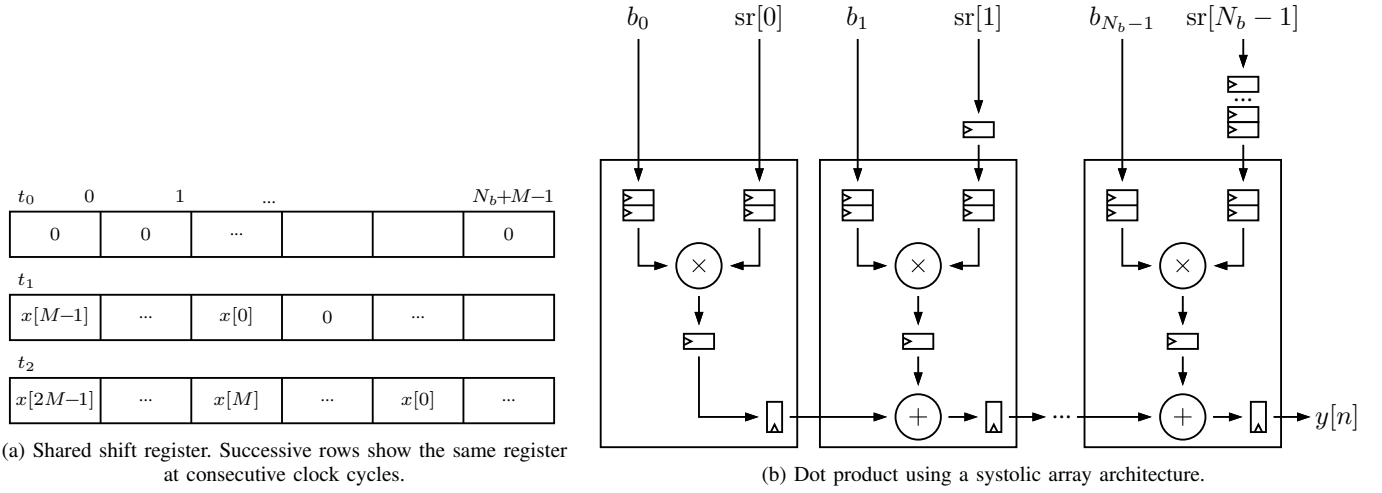
\centering
  \begin{subfigure}[b]{0.42\textwidth}\centering
    \includesvg[pretex=\scriptsize]{figures/shift_reg}
    \caption{Shared shift register. Successive rows show the same register at consecutive clock cycles.}\label{fig:shift_reg}
  \end{subfigure}\hfill
  \begin{subfigure}[b]{0.56\textwidth}\centering
    \includesvg{figures/dot_product_systolic}
    \caption{Dot product using a systolic array architecture.}\label{fig:dot_prod}
  \end{subfigure}
  \caption{Realization of the \(M\)-parallel FIR. \(\mathrm{sr}[k]\) indicates position \(k\) of the portion of the shift register seen by the dot product. The boundary boxes indicate DSP48E2 elements. The external registers are inferred by Vivado as Shift Register Logic (SRL) to save FPGA area.}\label{fig:fir}
\end{figure*}

A general \(N_{b}\)-tap FIR filter performs the convolution
\begin{equation}\label{eq:fir}
  y[n] = \sum_{k=0}^{N_{b}-1} b_{k}\, x[n-k],
\end{equation}
where the taps \(\{b_{0}, b_{1}, \ldots, b_{N_{b}-1}\}\) are also the filter's impulse response.
We implement a standard \(M\)-parallel FIR filter.
To relate \cref{eq:fir} to our \(M\)-parallel implementation, we make the variable substitution \(n = M m + r\), where \(m\) is the block index and \(r = 0, 1, \ldots, M-1\) indexes the parallel sub-filter.
Thus the \(r\)th sub-filter computes
\begin{equation}\label{eq:fir_sub_filter}
  y[M m + r] = \sum_{k=0}^{N_{b}-1} b_{k}\, x[M m + r - k].
\end{equation}
% At \(M = 2\), the \(r = 0\) sub-filter gives \(y[0] = b_{0} x[0]\) and \(y[2] = b_{0} x[2] + b_{1} x[1] + b_{2} x[0]\), and the \(r = 1\) sub-filter gives \(y[1] = b_{0} x[1]\) and \(y[3] = b_{0} x[3] + b_{1} x[2] + b_{2} x[1] + b_{3} x[0]\).
This is realized as a shared shift register (\cref{fig:shift_reg}) feeding \(M\) parallel FIR sub-filters, each implemented as a dot product.
Each dot product uses the same taps \(\{b_{0}, b_{1}, \ldots, b_{N_{b}-1}\}\) as one input array and the samples from a fixed position of the shift register as the other input array.
The sub-filter for \(r = 0\) extracts samples from shift register positions \(M-1\) to \(N_{b}+M-2\), that for \(r = 1\) from positions \(M-2\) to \(N_{b}+M-3\), and so on.
This arrangement directly implements \cref{eq:fir_sub_filter}.

This filter realizes the non-parametric residual correction of \cref{sec:residual}, and we also absorb impedance discontinuity reflections (\cref{sec:bounce}) into it.
Truncating \cref{eq:bounce} after the \(k = K\) term leaves a residual echo of relative amplitude \(\abs{\alpha_{e}}^{K+1}\), so the correction is adequate to the extent that \(K D + 1\) taps fit within the filter length.
For reflections whose correction extends beyond the reach of the FIR, a dedicated recursive section is a natural future addition.
The HDAWG's bounce filter, for comparison, accepts delays up to \SI{100}{ns}~\cite[Table~5.45]{zihdawg}.
% The dedicated recursive section would be \(y[n] = x[n] - \alpha_{e}\, y[n-D]\), which costs one multiplier instead of \(K\).
% It does not require the scattered look-ahead transformation when \(D \geq L M\), with \(L\) as defined in the IIR subsection, since its feedback delay already exceeds the pipeline depth.

\subsection{IIR}

Consider the general IIR difference equation
\begin{equation}\label{eq:iir}
  y[n] = \sum_{k=0}^{N_{b}-1} b_{k}\, x[n-k] + \sum_{k=1}^{N_{a}} a_{k}\, y[n-k].
\end{equation}
The dependency on \(y[n-1]\) spans a single sample period, which makes a naive implementation impossible to pipeline at full throughput: each output would have to complete an entire multiply-add-round-saturate within one clock cycle, whereas a DSP slice needs several cycles of latency to perform that operation at its maximum clock rate.
Full pipelining therefore requires breaking the single-cycle dependency.
We do this with scattered look-ahead pipelining~\cite{parhi1989pipeline}.
Clustered look-ahead~\cite{loomis1984highspeed, parhi1989pipeline, lim1992pipelined} transforms \cref{eq:iir} into \(y[n] = a'_{1} y[n-J] + a'_{2} y[n-J-1] + \ldots\) for a feedback latency of \(J\), whose feedback terms fall on different sample phases and so couple the branches of a super-sample-rate implementation; scattered look-ahead instead yields \(y[n] = a'_{1} y[n-J] + a'_{2} y[n-2J] + \ldots\), whose feedback terms all share the phase of \(y[n]\) when \(J\) is a multiple of \(M\).
In terms of the IIR's transfer function
\begin{equation}\label{eq:iir_h}
  H(z) = \frac{B(z)}{A(z)}
       = \frac{\sum_{k=0}^{N_{b}-1} b_{k} z^{-k}}{1 - \sum_{k=1}^{N_{a}} a_{k} z^{-k}},
\end{equation}
the transformation multiplies the numerator and denominator by a polynomial \(\Phi_{J}(z)\) chosen so that the denominator contains only the terms \(z^{-J}, z^{-2J}, \ldots, z^{-N_{a} J}\).
This is achieved with
\begin{equation}\label{eq:phi}
  \Phi_{J} (z) = \prod_{k=1}^{J-1} A \left(z e^{j 2 \pi k / J}\right).
\end{equation}
The cost of the scattered look-ahead transformation is a growth in the number of feedforward taps, from \(N_{b}\) to
\begin{equation}\label{eq:nb_prime}
  N_{b}' \coloneq N_{b} + N_{a} (J - 1).
\end{equation}
Here and below, a prime marks a quantity after the scattered look-ahead transformation.

The integrator uses a feedback loop latency of 2 clock cycles: one register holds the accumulation result and another holds the result of saturation.
The FOS and SOS use a feedback loop latency of 4 clock cycles, three of which lie inside the DSP slice: the first registers \(y\) and the coefficient value, the second holds the result of the multiplication, the third holds the result of the accumulation, and the fourth, in fabric, holds the result of rounding and saturation.

\begin{figure}\centering
  \includesvg{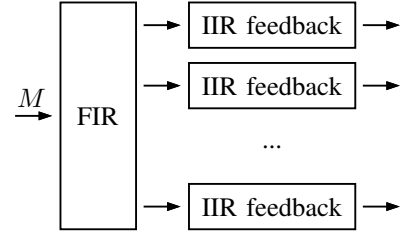}
  \caption{Super-sample-rate IIR architecture using \(M\) parallel IIR feedback sections.}\label{fig:iir}
\end{figure}

To make this work for a super-sample-rate architecture, \(M > 1\), we set the total look-ahead to \(J = L M\)~\cite{parhi1989block}, where \(L\) is the per-branch feedback pipeline depth required to achieve the maximum clock rate in a multiply-add-round-saturate operation.
Because \(J\) is a multiple of \(M\), every transformed feedback tap lands on the same polyphase branch as the output it feeds, so no feedback crosses between branches.
The resulting super-sample-rate IIR, shown in \cref{fig:iir}, reuses the existing super-sample-rate FIR for the feedforward section and adds \(M\) parallel feedback sections, each with an intrinsic feedback latency of \(L\) and each operating on one of the \(M\) samples per clock cycle.
Our polyphase feedback sections perform optional rounding and saturation after the full feedback sum.
The loop latencies above are \(L = 2\) for the integrator and \(L = 4\) for the FOS and SOS, so at \(M = 2\) the integrator uses \(J = 4\) and the sections use \(J = 8\).

The factors of \(A(z) \Phi_{J}(z)\) group by root into \(\prod_{k=0}^{J-1} \left(1 - p_{i} e^{-j 2 \pi k / J} z^{-1}\right) = 1 - p_{i}^{J} z^{-J}\), with \(p_{i}\) the roots of \(A(z)\), so the general form of \(A'(z)\) is
\begin{equation}
  A'(z) = \prod_{k=1}^{N_{a}} \left(1 - p_{k}^{J} z^{-J}\right).
\end{equation}
Expanding the product relates the transformed coefficients to the poles:
\begin{equation}
  a_{k}' = (-1)^{k+1} e_{k} \left(p_{1}^{J}, p_{2}^{J}, \ldots, p_{N_{a}}^{J}\right),
\end{equation}
where
\begin{equation*}
  e_{k} \left(X_{1}, X_{2}, \ldots, X_{n}\right) = \sum_{1 \leq c_{1} < c_{2} < \cdots < c_{k} \leq n} X_{c_{1}} X_{c_{2}} \cdots X_{c_{k}}
\end{equation*}
is the elementary symmetric polynomial.

In the case of the high-pass droop compensation filter described by \cref{eq:integrator_transfer_function},
\begin{equation}\label{eq:high_pass_aprime}
  A'(z) = 1 - z^{-J},
\end{equation}
and, using the identity \(\Phi_{J}(z) = A'(z) / A(z)\),
% \(B'(z) = B(z) \Phi_{J}(z) = \left(1 + b_{1} z^{-1}\right) \left(1 - z^{-J}\right) / \left(1 - z^{-1}\right) = \left(1 + b_{1} z^{-1}\right) \sum_{k=0}^{J-1} z^{-k}\).
% Expanding gives \(1 + \sum_{k=1}^{J-1} z^{-k} + b_{1} \sum_{k=1}^{J} z^{-k}\), which collects to \cref{eq:high_pass_bprime}.
\begin{equation}\label{eq:high_pass_bprime}
  B'(z) = 1 + \left(1 + b_{1}\right) \sum_{k=1}^{J-1} z^{-k} + b_{1} z^{-J}.
\end{equation}

The exponential tail compensation filter of \cref{eq:tail_pre} has the single pole \(p_{1}\), so \(\Phi_{J}(z) = \left(1 - p_{1}^{J} z^{-J}\right) / \left(1 - p_{1} z^{-1}\right) = \sum_{k=0}^{J-1} p_{1}^{k} z^{-k}\), and
% \(A'(z) = 1 - p_{1}^{J} z^{-J}\) is the \(N_{a} = 1\) case of the general form above.
% Then \(B'(z) = \kappa \left(1 - q_{1} z^{-1}\right) \sum_{k=0}^{J-1} p_{1}^{k} z^{-k} = \kappa \sum_{k=0}^{J-1} p_{1}^{k} z^{-k} - \kappa q_{1} \sum_{k=1}^{J} p_{1}^{k-1} z^{-k}\).
% The \(k = 0\) term is \(\kappa\), the \(k = J\) term is \(-\kappa q_{1} p_{1}^{J-1}\), and for \(1 \leq k \leq J-1\) the two sums combine to \(\kappa p_{1}^{k-1} \left(p_{1} - q_{1}\right)\).
\begin{subequations}\label{eq:tail_transformed}
\begin{align}
  A'(z) &= 1 - p_{1}^{J} z^{-J},\\
  \begin{split}
    B'(z) &= \kappa \Biggl[1
             + \left(p_{1} - q_{1}\right) \sum_{k=1}^{J-1} p_{1}^{k-1} z^{-k}\\
          &\qquad - q_{1} p_{1}^{J-1} z^{-J}\Biggr].
  \end{split}
\end{align}
\end{subequations}
The high-pass droop is the \(\kappa = 1\), \(p_{1} = 1\) case.

The damped oscillation compensation filter of \cref{eq:sos} has two poles, so its \(\Phi_{J}(z)\) is a product of two factors of the single-pole form above, and the \(z^{-m}\) coefficient of that product is a convolution of the truncated geometric sequences \(p_{1}^{k}\) and \(p_{2}^{k}\), giving
% \(A'(z) = \prod_{l=1}^{2} \left(1 - p_{l}^{J} z^{-J}\right)\), expanded below.
% \(\Phi_{J}(z) = A'(z) / A(z) = \prod_{l=1}^{2} \sum_{k=0}^{J-1} p_{l}^{k} z^{-k}\), so its \(z^{-m}\) coefficient \(\varphi_{m}\) is the convolution \(\sum_{k} p_{1}^{k} p_{2}^{m-k}\) taken over \(\max(0, m-J+1) \leq k \leq \min(m, J-1)\).
% For \(m \leq J-1\) that range is \(0 \leq k \leq m\), which sums to \(h_{m}\).
% For \(m \geq J-1\) it is \(m-J+1 \leq k \leq J-1\), and factoring out \(\left(p_{1} p_{2}\right)^{m-J+1}\) leaves \(h_{2J-2-m}\); the two agree at \(m = J-1\).
% Multiplying \(\Phi_{J}(z)\) by \(B(z) = \kappa \left(1 - q_{1} z^{-1}\right) \left(1 - q_{2} z^{-1}\right)\) then gives the three-term convolution below.
% In closed form \(h_{m} = \left(p_{1}^{m+1} - p_{2}^{m+1}\right) / \left(p_{1} - p_{2}\right)\), which is singular at \(p_{1} = p_{2}\) and complex termwise for a conjugate pair, so we generate it by its recursion instead.
\begin{subequations}\label{eq:osc_transformed}
\begin{align}
  A'(z) &= 1 - \left(p_{1}^{J} + p_{2}^{J}\right) z^{-J}
           + \left(p_{1} p_{2}\right)^{J} z^{-2J},\\
  \begin{split}
    B'(z) &= \kappa \sum_{m=0}^{2J}
             [\varphi_{m}
              - \left(q_{1} + q_{2}\right) \varphi_{m-1}\\
          &\qquad\qquad + q_{1} q_{2}\, \varphi_{m-2}] z^{-m},
  \end{split}
\end{align}
\end{subequations}
where \(\varphi_{m} \coloneq 0\) outside \(0 \leq m \leq 2J-2\) and
\begin{equation}\label{eq:osc_phi_coeff}
  \varphi_{m} = \left(p_{1} p_{2}\right)^{\max(0,\, m-J+1)}\,
                h_{\min(m,\, 2J-2-m)},
\end{equation}
with \(h_{m}\) the complete homogeneous symmetric polynomial in \(p_{1}\) and \(p_{2}\), generated by
\begin{equation}\label{eq:osc_h_recursion}
  \begin{split}
    h_{m} &= \left(p_{1} + p_{2}\right) h_{m-1} - p_{1} p_{2}\, h_{m-2},\\
    h_{0} &= 1, \quad h_{-1} = 0.
  \end{split}
\end{equation}
The same recursion supplies the feedback coefficient \(p_{1}^{J} + p_{2}^{J} = h_{J} - p_{1} p_{2}\, h_{J-2}\).
\Cref{eq:sos} gives \(q_{1} + q_{2} = 2 \rho \cos\theta\) and \(q_{1} q_{2} = \rho^{2}\), and \cref{eq:osc_z} gives \(p_{1} + p_{2} = -c_{1}/c_{0}\) and \(p_{1} p_{2} = c_{2}/c_{0}\), so \cref{eq:osc_transformed} is real and does not require root finding.

\subsection{DSP48E2 Mapping}

The dot products themselves use a systolic array architecture~\cite{kung1979systolic}, shown in \cref{fig:dot_prod}, which is efficiently supported by the Xilinx DSP48E2 and requires minimal external fabric logic.
This architecture uses the built-in multiply-accumulate capabilities of the DSP slice as well as the dedicated PCIN/PCOUT cascade that routes directly between slices in a column to avoid the use of general fabric routing.
Finally, we use the built-in pattern detect circuitry of the DSP slice~\cite{ug579, vivadoSynthesis}, which allows us to perform convergent rounding in the final DSP slice.
The pattern detect circuitry can also perform saturation, but not simultaneously with rounding, so when a module instance enables both, rounding uses the pattern detect and saturation is performed in fabric following the final slice.

The feedback sections reuse the same slice cascade.
Each of the \(M\) polyphase feedback sections is a systolic cascade of \(N_{a}\) slices, one per transformed feedback tap; the feedforward output enters through the C port of the first slice, and the last slice applies convergent rounding through its pattern detect.

\section{Fixed-Point Design}\label{sec:fixed_point}

The correction filters must correct the line's step response to within a specified error over the expected physical parameter range of \cref{sec:distortion}.
We therefore relate step response error to physical parameter and transfer function tolerances, and those tolerances, together with the parameter ranges, to the transformed coefficient and datapath Q formats.
A maximum step response error, chosen from two-qubit gate fidelity requirements, then fixes the Q formats.
We must additionally ensure that the quantized poles stay inside the unit circle to maintain stability of the realized filter.
Throughout, \(\Delta X\) denotes the quantization step (one least significant bit, LSB) of a quantity \(X\), \(\hat{X}\) the value realized in fixed point, and \(\delta X \coloneq \hat{X} - X\) the change from quantization.
We additionally extend the hat notation to polynomials formed from quantized coefficients to designate design-realized polynomials.
Within a filter, all feedback coefficients share one Q format and all feedforward coefficients share another, so \(\Delta a'\) and \(\Delta b'\) each denote a single step.
We round to nearest, so a stored coefficient satisfies \(\abs{\delta X} \leq \Delta X / 2\); for quantities derived from the coefficients, \(\delta X\) is taken to first order in those coefficient errors.
\Cref{tab:fixed_point} collects the results.

The correction is realized as \(\hat{H}\), so the corrected line has response \(G \hat{H} = 1 + \mathcal{F}\) with \(\mathcal{F} \coloneq \hat{H} / H - 1\), and the step response error of the corrected line is the step response of \(\mathcal{F}\).
% \(\hat{H} = 1 / (G + \delta G)\) gives \(\delta H = -\delta G\, H^{2}\) to first order, and \(\mathcal{F} = \delta H / H\).
To first order, an error \(\delta x\) in a physical parameter \(x\) of the correction gives \(\mathcal{F} = -\delta G / G\), the fractional error of the model, with \(\delta G \coloneq (\partial G / \partial x)\, \delta x\).
Most prior work targets a step response error after predistortion of approximately \(0.1\%\)~\cite{rol2020cryoscope, guo2024universal}, which we target as well and denote \(t_{\mathrm{step}}\).
We require the peak step response error attributed to each bounded quantity to be at most \(t_{\mathrm{step}}\) in its own worst case, without summing the contributions, since those worst cases generally fall at different parameters, times, and rounding patterns.

The bounds in \cref{tab:fixed_point} follow from the sensitivity of the realized poles and zeros to coefficient quantization\longonly{, derived in \cref{sec:stability,sec:pole_sensitivity,sec:zero_sensitivity,sec:param_sensitivity,sec:cancellation}}.
\(\mathcal{B}_{p}\) and \(\mathcal{B}_{z}\) bound the fractional displacements \(\abs{\delta p_{i} / p_{i}}\) and \(\abs{\delta q_{i} / q_{i}}\) of the poles and zeros of \(H\) under round to nearest.
\(\mathcal{E}\) bounds the displacement \(\abs{\delta u + j \delta v}\) of the oscillation residue, with \(u \coloneq \alpha_{r} \cos\phi\) and \(v \coloneq \alpha_{r} \sin\phi\).
The \(N_{a}(J-1)\) zeros that \(\Phi_{J}\) adds cancel the added poles only in exact arithmetic~\cite[Sec.~IV-B]{parhi1989pipeline}; quantizing the feedforward coefficients displaces them from the poles~\cite[Sec.~II-B]{parhi1991finite}.
\(E(z)\) denotes the fractional error of the realized transfer function relative to the transfer function with quantized poles and exact zeros, and \(\lVert E \rVert_{\infty}\) its maximum over frequency.
To limit it, we form \(B'\) as \(B \hat{\Phi}_{J}\), with \(\hat{\Phi}_{J}\) built from the quantized poles, before quantizing its coefficients.
\shortonly{The derivations and the explicit forms of these bounds are given in the extended version of this paper~\cite{huszabianlou2026extended}.}

We check the step response error of the resulting design in simulation: a bit-accurate model of the filter datapath is swept over the parameter ranges of \cref{tab:fixed_point}, and the worst-case error over each range is reported in the simulated column.
For the IIR filters this is the peak step response error of the corrected line for a step of \num{0.5} of full scale, or \num{0.1} for the high-pass droop, whose output ramp would otherwise saturate at the shortest simulated \(\tau\).
The droop is simulated over an \SI{8}{\us} pulse, longer than any flux pulse we expect.
For the FIR it is the peak output error under random taps and full-scale random input, which includes the half LSB of the final round.
Building \(B'\) from the exact poles rather than the quantized ones raises the simulated damped oscillation error by nearly an order of magnitude.

\begin{table*}
  \centering
  \caption{%
    Results of \cref{sec:fixed_point} at \(J = 8\) (\(J=4\) for the integrator) and \(T_{s} = \SI{1}{ns}\).
    Dashes indicate columns not applicable to a particular filter.
    The accuracy column lists the peak step response error contributed by each bounded quantity, to first order in the coefficient errors, except that the \(\lVert E \rVert_{\infty}\) entries bound the fractional transfer function error.
  }\label{tab:fixed_point}
  \footnotesize
  \setlength{\tabcolsep}{4pt}
  \begin{tabularx}{\textwidth}{@{}l Y{0.7} Y{0.8} c c c c Y{1.25} Y{0.95} c@{}}
    \toprule
    & \multicolumn{2}{c}{constraints on} & \multicolumn{4}{c}{Q format} & & & \\
    \cmidrule(lr){2-3} \cmidrule(lr){4-7}
    filter & \(\Delta a'\) & \(\Delta b'\) & \(a'\) & \(b'\) & feedforward & accumulator
      & range covered & accuracy & simulated \\
    \midrule
    high-pass droop
      & unity tap, exact
      & \(\tau\)\longonly{, \labelcref{eq:integrator_tau_bound}}
      & --- & Q2.25 & Q2.29 & Q1.29
      & \(\tau \leq \SI{67}{\us}\)
      & \(\abs{\delta \tau / \tau} \leq 0.1\%\)
      & \(0.03\%\) \\
    \addlinespace
    exponential tail
      & \(\alpha\)\longonly{, \labelcref{eq:tail_alpha_bound}}
      & \(\alpha\), \(\lVert E \rVert_{\infty}\), \labelcref{eq:tail_tb}
      & Q1.17 & Q2.20 & Q2.22 & Q1.22
      & \(\abs{\alpha} \leq 0.4\), \(\SI{30}{ns} \leq \tau \leq \SI{500}{ns}\)
      & \(\abs{\delta \alpha} / (1 {+} \alpha) \leq 0.1\%\) \newline \(\lVert E \rVert_{\infty} \leq 0.1\%\)
      & \(0.04\%\) \\
    \addlinespace
    damped oscillation
      & stability, \(\mathcal{E}\)\longonly{, \labelcref{eq:osc_stability_bound,eq:osc_amp_phase_bound}}
      & \(\tau\), \(T_{\mathrm{osc}}\) jointly \newline \(\mathcal{E}\) \newline \(\lVert E \rVert_{\infty}\)\longonly{ \newline \labelcref{eq:osc_zero_bound,eq:osc_amp_phase_bound,eq:residual_bound}}
      & Q2.16 & Q3.24 & Q3.26 & Q1.26
      & \(\SI{5}{MHz} \leq 1/T_{\mathrm{osc}} \leq \SI{150}{MHz}\) (excluding \(\pm\SI{1.0}{MHz}\) about \SI{62.5}{MHz} and \SI{125}{MHz}) at \(Q \geq 3\), \(\SI{30}{ns} \leq \tau \leq \SI{300}{ns}\), \(\abs{\alpha_{r}} \leq 0.05\)
      & \(2 \mathcal{E} \leq 0.11\%\) \newline \(\tau\), \(T_{\mathrm{osc}} < 0.01\%\) \newline \(\lVert E \rVert_{\infty} \leq 0.11\%\)
      & \(0.04\%\) \\
    \addlinespace
    FIR
      & ---
      & output error, \labelcref{eq:fir_bits}
      & --- & Q3.20 & --- & ---
      & \(N_{b} = 20\), band-averaged \(\abs{H} \leq 4\)
      & output error below half an output LSB
      & \num{0.60} LSB \\
    \bottomrule
  \end{tabularx}
\end{table*}

\subsection{Q Format Selection}\label{sec:q_format}

We use a Q number format of Q\(I\).\(F\), where \(I\) is the number of integer bits, including the sign bit, and \(F\) is the number of fractional bits.
We subscript \(I\) and \(F\) to identify the signal: \(a\) and \(b\) denote the feedback and feedforward coefficients, respectively (e.g., \(\Delta a' = 2^{-F_{a}}\)), \(x\) denotes the filter input, and \(\mathrm{acc}\) denotes an IIR accumulator.
Often the same constraint bounds both \(\Delta a'\) and \(\Delta b'\).
In this case we introduce a parameter \(\lambda \in (0, 1)\) representing the share of the tolerance borne by the feedback term, leaving \(1 - \lambda\) for the feedforward term.
The RFSoC DAC uses a 16-bit datapath~\cite{pg269}, which we arbitrarily designate as Q1.15.
The input and output of each filter in the cascade use a matching Q1.15 format.
The high-pass droop, exponential tail, and damped oscillation corrections below are realized by the integrator, the FOS, and the SOS, respectively.

% The restriction on \(\omega\) is needed because \(\abs{B'}\) vanishes wherever an added zero is on the unit circle, as it does for the high-pass, and there \(A'\) vanishes with it under \cref{eq:high_pass_restrict_perturbation}.
% \Cref{sec:cancellation} evaluates this minimum in \cref{eq:tail_residual} and, for the oscillation, directly; the high-pass takes \(B'(1)\) from \cref{eq:high_pass_bprime}.
% Round to nearest passes an input LSB whose image exceeds half an accumulator LSB, so requiring a full LSB is conservative by one bit.
The accumulator fractional width follows from the smallest input amplitude that must survive rounding into the accumulator.
One input LSB at frequency \(\omega\) reaches the accumulator as \(2^{-F_{x}} \abs{B'(e^{j \omega})}\), which survives when
\begin{equation}\label{eq:facc}
  F_{\mathrm{acc}} \geq F_{x} - \log_{2} \left(\min_{\omega} \abs{B'(e^{j \omega})}\right),
\end{equation}
where \(\omega\) is restricted to the band corrected by the filter.
We do not scale between the accumulator and output, so \(I_{\mathrm{acc}} = I_{x} = 1\).
The feedforward word entering the recursion is \(B' x = \hat{A}' y\), however, so it is bounded by \(\lVert \hat{A}' \rVert_{\infty}\) times the accumulator range and carries its own integer width.

\subsubsection{High-Pass Droop}\label{sec:q_high_pass}

Quantizing \(b_{1}'\) displaces only the zero, so the step response error is \(-(\delta \tau / \tau) (1 - e^{-t/\tau})\) to first order, whose magnitude rises monotonically to \(\abs{\delta \tau / \tau}\)\longonly{; \(E\) of \cref{eq:integrator_residual} is this same error, since the pole is exact}.
Over a pulse of duration \(T_{p} \ll \tau\) the error reaches only the fraction \(T_{p} / \tau\) of this value, so the requirement below is conservative for such pulses.
Requiring \(\abs{\delta \tau / \tau} \leq t_{\mathrm{step}}\)\longonly{ in \cref{eq:integrator_tau_bound}} gives
\begin{equation}\label{eq:integrator_bits}
  F_{b} = -\log_{2} \left\{2 \rho \left[\exp \left(\frac{T_{s}}{\tau}
            \frac{t_{\mathrm{step}}}{1 + t_{\mathrm{step}}}\right) - 1\right]\right\}.
\end{equation}
% \(\exp x - 1 \to x\), \(\rho \to 1\), and \(T_{s} / \tau \to 1 - \rho\).
For \(T_{s} \ll \tau\) and \(t_{\mathrm{step}} \ll 1\) this reduces to \(F_{b} = \log_{2} \left[1 / \left(2 (1 - \rho) t_{\mathrm{step}}\right)\right]\).
Our bias tees have a nominal \(\tau = \SI{18}{\us}\), which at \(t_{\mathrm{step}} = 0.1\%\) requires 23.1 fractional bits.
Bias tees reported elsewhere have \(\tau\) of \SI{13}{\us}~\cite[Fig.~2]{aggarwal2025transients} and \SI{19.2}{\us}~\cite[App.~D]{hellings2025calibrating}.
\Cref{eq:high_pass_bprime} shows that 2 integer bits are needed, including the sign bit, since the \(b_{0}' = +1\) coefficient must be representable.
% \(F_{b} = 25\) and \(t_{\mathrm{step}} = 0.1\%\) give \(1 - \rho = 2^{-25} / 0.002 = \num{1.49e-5}\), and \(\tau = -T_{s} / \ln \rho = \SI{67.1}{\us}\).
% Equivalently, 25 exceeds the 23.10 bits of the \SI{18}{\us} case by 1.90, and \(18 \times 2^{1.90} = 67\).
We settle on a Q format of Q2.25, which allows us to represent time constants up to \SI{67}{\us} at \(t_{\mathrm{step}} = 0.1\%\) and costs essentially the same FPGA resources as Q2.23 or Q2.24, since the DSP48E2 provides a \(27 \times 18\)-bit multiplier and the inputs are 16 bits wide.

% \(\abs{B'}^{2} = \left(1 - 2 \rho \cos\omega + \rho^{2}\right) \sin^{2}(J \omega / 2) / \sin^{2}(\omega / 2)\) exceeds its \(\omega = 0\) value \(J^{2} (1 - \rho)^{2}\) wherever \(\abs{\sin (J \omega / 2)} \geq J (1 - \rho) / (2 \sqrt{\rho})\), which at \(\tau \leq \SI{67}{\us}\) and \(J = 4\) excludes only \SI{2.4}{kHz} on either side of each null.
For the accumulator, \(\Phi_{J}\) places zeros of \(B'\) at multiples of \(f_{s} / J\), all above the corrected band, and over the corrected band \(\abs{B'(e^{j \omega})}\) is minimized at DC, where \cref{eq:high_pass_bprime} gives \(B'(1) = J (1 - \rho)\).
With \(\rho\) restricted through \(\tau \leq \SI{67}{\us}\), \cref{eq:facc} then requires \(F_{\mathrm{acc}} \geq 29\), so we use Q1.29.

\subsubsection{Exponential Tail}

% \(\delta \alpha\) from \(\delta q_{1}\) alone is \(-\delta q_{1} / \left[\kappa (1 - q_{1})\right]\) by \cref{eq:tail_param_sensitivity}, so its step contribution \(\abs{\delta \alpha} / (1 + \alpha) = \abs{\delta q_{1}} / (1 - q_{1}) \to (\tau / T_{s}) \abs{\delta \rho / \rho}\), while \(\tau\) contributes \((\abs{\alpha} / e) \abs{\delta \tau / \tau} \to (\abs{\alpha} / e) (\tau / T_{s}) \abs{\delta \rho / \rho}\) by \cref{eq:tau_from_zero}; the ratio is \(\abs{\alpha} / e\), at most \(0.4 / e = 0.15\), for every \(\lambda\).
% Tail: \(G(s) = (1 + s \tau_{2}) / (1 + s \tau)\) with \(\tau_{2} \coloneq (1 + \alpha) \tau\), so \(\partial G / \partial \tau = \alpha s / (1 + s \tau)^{2}\) and \(\partial G / \partial \alpha = s \tau / (1 + s \tau)\).
% For \(\tau\), \(\mathcal{F} = -\alpha (\delta \tau / \tau)\, s \tau / \left[(1 + s \tau)(1 + s \tau_{2})\right]\), whose step response is \(-(\delta \tau / \tau) \left(e^{-t/\tau_{2}} - e^{-t/\tau}\right)\) by partial fractions, stationary at \(t / \tau = (1 + \alpha) \ln (1 + \alpha) / \alpha\) with value \(-\alpha (1 + \alpha)^{-(1 + \alpha)/\alpha}\, \delta \tau / \tau\); the factor \((1 + \alpha)^{-(1 + \alpha)/\alpha}\) is \(1/e\) as \(\alpha \to 0\) and within 26\% of it over \(\abs{\alpha} \leq 0.4\).
% For \(\alpha\), \(\mathcal{F} = -\left[\delta \alpha / (1 + \alpha)\right] s \tau_{2} / (1 + s \tau_{2})\), whose step response \(-\left[\delta \alpha / (1 + \alpha)\right] e^{-t/\tau_{2}}\) peaks at \(t = 0\); in discrete time the \(n = 0\) value is \(\delta \kappa / \kappa\).
With \(\tau_{2} \coloneq (1 + \alpha) \tau\), the step response errors are \(-(\delta \tau / \tau) (e^{-t/\tau_{2}} - e^{-t/\tau})\) from \(\tau\) and \(-\left[\delta \alpha / (1 + \alpha)\right] e^{-t/\tau_{2}}\) from \(\alpha\), to first order, with peaks of \(\abs{\alpha} \abs{\delta \tau / \tau} / e\) at \(t = \tau\) for small \(\abs{\alpha}\) and \(\abs{\delta \alpha} / (1 + \alpha)\) at \(t = 0\).
We therefore require \(\abs{\delta \alpha} \leq (1 + \alpha) t_{\mathrm{step}}\)\longonly{ in \cref{eq:tail_alpha_bound}}.
% \(1 / B' = G / (1 - p_{1}^{J} z^{-J})\), the step response of \(G\) is \(g[n] = 1 + \alpha \rho^{n} > 0\), and the impulse response of \(1 / (1 - p_{1}^{J} z^{-J})\) is \(p_{1}^{J m} \geq 0\) at \(n = J m\) for \(\alpha \geq -\rho\), so the step response of \(1 / B'\) is nonnegative and at most \(\left[1 + \max(\alpha, 0)\right] / (1 - p_{1}^{J})\).
% The step response of \cref{eq:residual} is then at most \((J + 1) (\Delta b' / 2) \left[1 + \max(\alpha, 0)\right] / (1 - p_{1}^{J})\), which is \cref{eq:tail_residual} times \(\left[1 + \max(\alpha, 0)\right] (1 + \rho) / \left[1 + \rho + 2 \max(\alpha, 0)\right] \leq 1\), with equality for \(\alpha \leq 0\).
We also require \(\lVert E \rVert_{\infty} \leq t_{\mathrm{step}}\)\longonly{, which bounds the peak step response error from \(E\) by \cref{eq:tail_residual}, because the step response of \(1 / B'\) is nonnegative}.
% The stability constraint does not govern: for \(\alpha > -\rho\), \(1 - p_{1}^{J} = \kappa (1 - q_{1}) \sum_{k=0}^{J-1} p_{1}^{k} \geq J \kappa (1 - q_{1}) p_{1}^{J-1}\), so \cref{eq:exp_stability} permits a \(\Delta a'\) at least \(1 / (\lambda t_{\mathrm{step}})\) times that of \cref{eq:tail_alpha_bound}, exactly and without expanding \(1 - p_{1}^{J}\).
% The \(\lambda\) share of \cref{eq:tail_alpha_bound} is \(\abs{p_{1}} \mathcal{B}_{p} / \kappa \leq \lambda \kappa (1 - q_{1}) (1 + \alpha) t_{\mathrm{step}}\), which \cref{eq:exp_pole_bound} inverts to \(\Delta a' \leq 2 J \lambda \kappa^{2} \abs{p_{1}}^{J-1} (1 - q_{1}) (1 + \alpha) t_{\mathrm{step}}\).
% The \(1 - \lambda\) share of \cref{eq:tail_alpha_bound} requires \(q_{1} \mathcal{B}_{z} \leq (1 - \lambda) \kappa (1 - q_{1}) (1 + \alpha) t_{\mathrm{step}}\), constraining \(\mathcal{B}_{z}\); \cref{eq:tail_zero_bound} carries it to \(\Delta b'\), which \cref{eq:tail_residual} constrains directly.
% Reducing: \(\sum_{k=0}^{J} \rho^{-k} \to J+1\), \(\sum_{k=0}^{J-1} (p_{1}/\rho)^{k} \to J\), \(\exp x - 1 \to x\), \(1 - \rho \to T_{s}/\tau\), \(\rho \to 1\), and \(\abs{p_{1}} \to 1\).
Assuming \(J T_{s} \ll (1 + \alpha) \tau\), these imply
\begin{equation}\label{eq:tail_bits}
  \begin{split}
    F_{a} &= \log_{2} \frac{\left(1 + \alpha\right) \tau}
                           {2 J \lambda T_{s} t_{\mathrm{step}}},\\
    F_{b} &= \log_{2} \frac{\left(J + 1\right) \left(1 + \alpha\right) \tau}
                           {2 J T_{s} t_{b}},
  \end{split}
\end{equation}
where
\begin{equation}\label{eq:tail_tb}
  t_{b} \coloneq \min \left\{\frac{t_{\mathrm{step}}}{1 + \max(\alpha, 0)},\,
                             \left(1 - \lambda\right) t_{\mathrm{step}}\right\}.
\end{equation}
% \(\abs{p_{1}}^{J-1} (1 - \rho) / (T_{s}/\tau) \approx 1 - J T_{s} / \left[(1+\alpha) \tau\right]\), so \cref{eq:tail_bits} overstates the \(\Delta a'\) limit by that factor.
% On the \(\Delta b'\) side \(\sum \rho^{-k} / (J+1) \approx 1 + J T_{s} / (2 \tau)\) and \(J / \sum (p_{1}/\rho)^{k} \approx 1 - J \alpha T_{s} / \left[2 (1+\alpha) \tau\right]\) partly cancel, leaving \(1 + J T_{s} / \left[2 (1+\alpha) \tau\right]\).
% Taking \(\log_{2}\) of each.

% \(b'_{0} = \kappa\), \(b'_{k} = \kappa p_{1}^{k-1} (p_{1} - q_{1})\) for \(1 \leq k \leq J-1\), and \(b'_{J} = -\kappa q_{1} p_{1}^{J-1}\), with \(\abs{p_{1} - q_{1}} = \abs{\alpha} \kappa (1 - q_{1}) \ll 1\), so \(\abs{b_{0}'}\) is the largest.
Stability constrains \(\abs{a_{1}'} < 1\), and \(\max_{k} \abs{b_{k}'} = 1 / (1 + \alpha)\), so \(I_{a} = 1\) and \(I_{b}\) is determined by the most negative value of \(\alpha\).

% The minimum is the denominator of \cref{eq:tail_residual} without its \(J+1\), and is exact for even \(J\), since \(\abs{H}\) is least at \(\omega = \pi\) for \(\alpha > 0\) and at \(\omega = 0\) otherwise, and \(\omega = \pi\) is then also an added zero.
% \(B'(1) = B(1) \Phi_{J}(1) = \kappa (1 - q_{1}) (1 - p_{1}^{J}) / (1 - p_{1}) = 1 - p_{1}^{J}\), since \(1 - p_{1} = \kappa (1 - q_{1})\); this also confirms unity DC gain, the steady state being \(x_{0} B'(1) / (1 - a_{1}') = x_{0}\).
% \(1 - p_{1}^{J} \approx J (1 - p_{1})\) recovers the integrator's \(J (1 - \rho)\) as \(\alpha \to 0\), and \(\rho \to 1\) sends the \(\max(\alpha, 0)\) factor to \(1 + \max(\alpha, 0)\).
For the accumulator, \longonly{\cref{eq:tail_residual} makes }\(\min_{\omega} \abs{B'} = \left(1 + \rho\right) \left(1 - p_{1}^{J}\right) / \left[1 + \rho + 2 \max(\alpha, 0)\right]\) for even \(J\), so \cref{eq:facc} requires \(F_{\mathrm{acc}} \geq F_{x} + \log_{2} \left\{\left[1 + \rho + 2 \max(\alpha, 0)\right] / \left[\left(1 + \rho\right) \left(1 - p_{1}^{J}\right)\right]\right\}\), or \(F_{x} + \log_{2} \left\{\left(1 + \alpha\right) \left[1 + \max(\alpha, 0)\right] \tau / (J T_{s})\right\}\) under the same \(J T_{s} \ll (1 + \alpha) \tau\) assumption.

% The worst case is \(\tau = \SI{500}{ns}\) and \(\alpha = +0.4\), where the \((1 - \lambda) t_{\mathrm{step}}\) entry of \cref{eq:tail_tb} governs, at \(\num{5e-4}\) against \(\num{7.1e-4}\) for \(\lVert E \rVert_{\infty}\).
% There the reduced \cref{eq:tail_bits} gives 16.42 and 19.59 and the exact \cref{eq:tail_bits_exact} gives 16.43 and 19.59.
% \(\max_{k} \abs{b_{k}'} = 1/(1+\alpha)\) reaches \(5/3\) at \(\alpha = -0.4\), so \(I_{b} = 2\).
The \SI{30}{ns} floor follows the division of prior calibrations, which assign dynamics above \SIrange{25}{30}{ns} to IIR sections and the remainder to the FIR~\cite{rol2020cryoscope, hellings2025calibrating}.
Tails reported between this floor and \SI{500}{ns} have amplitudes of about 0.02 or less~\cite[Table~VI]{li2025couplerpulse}, \cite[Table~V]{li2024doubletransmon}, \cite[Sec.~4]{foxen2019fluxsampling}, except for one of amplitude 0.39 at \SI{36}{ns}~\cite[Table~III]{glaser2024sensitivity}; shorter tails, with amplitudes up to 0.6~\cite[Table~III]{glaser2024sensitivity}, \cite[Sec.~4]{foxen2019fluxsampling}, are left to the FIR.
Over \(\abs{\alpha} \leq 0.4\) and \(\SI{30}{ns} \leq \tau \leq \SI{500}{ns}\), at \(t_{\mathrm{step}} = 0.1\%\) and \(\lambda = 1/2\), \cref{eq:tail_bits,eq:facc} require \(F_{a} = 16.4\), \(F_{b} = 19.6\), and \(F_{\mathrm{acc}} = 21.9\), so \(a'\), \(b'\), and the accumulator use Q1.17, Q2.20, and Q1.22.
Tails of \SIrange{0.5}{1.3}{\us} with amplitudes of 0.02 or less have also been reported~\cite[Table~VI]{li2025couplerpulse}, \cite[Table~V]{li2024doubletransmon}, \cite[Table~III]{glaser2024sensitivity}; \cref{eq:tail_bits} shows that at fixed \(F_{a}\) and \(F_{b}\) the step response error for such tails rises in proportion to \(\tau\).

\subsubsection{Damped Oscillation}\label{sec:q_osc}

Approximating \(H\) by unity in \(\mathcal{F}\), the step response error from \(\delta \tau\) and \(\delta T_{\mathrm{osc}}\) peaks at \(t = \tau\), where it is at most \((2 \abs{\alpha_{r}} / e) (\tau / T_{s}) \mathcal{B}_{z}\), and the residue displacement produces a peak of \(2 \mathcal{E}\) at \(t = 0\)\longonly{ (\cref{eq:osc_step_error})}.
We require each to be at most \(t_{\mathrm{step}}\)\longonly{, \cref{eq:osc_step_tolerances}}, and \(\lVert E \rVert_{\infty} \leq t_{\mathrm{step}}\).
The bounds on \(\mathcal{B}_{p}\), \(\mathcal{B}_{z}\), and \(\mathcal{E}\) diverge for a resonance near DC or Nyquist and where the feedback delay \(J T_{s}\) spans a multiple of a half period of the poles or of the mean of the poles and zeros\longonly{ (\cref{sec:q_format_derivations})}, so no finite width covers the whole parameter space.
We instead fix the formats from the available datapath and state the range of physical parameters they correct and to what accuracy.

% TODO The following section would benefit from being rewritten in a more rigorous and less heuristic way. Unfortunately, I've found this extremely hard to do.

In the feedback section, the coefficient and accumulator compete for the 27- and 18-bit DSP48E2 ports.
% \cref{eq:facc} reads \(\min_{\omega} \abs{B'} \geq 2^{F_{x} - F_{\mathrm{acc}}}\), which is \(2^{-2}\) at 17 fractional bits and \(2^{-11}\) at 26.
% Swept at \(\alpha_{r} = 0\) with \(\tau \geq \SI{30}{ns}\), the largest \(Q\) that \(2^{-2}\) permits is zero at \SI{5}{MHz}, below 3 at \SI{20}{MHz}, and only in the tens at \SI{150}{MHz}, while \(2^{-11}\) permits \(Q\) above 100 throughout.
Restricting the accumulator to 17 fractional bits, of which 15 are already taken by \(F_{x}\), would overly limit the range of distortions we can correct, so we give it the 27-bit port.
Therefore, the accumulator uses Q1.26 and \(a'\) is left with the 18-bit port.

With \(p_{1,2} = \rho_{p} e^{\pm j \theta_{p}}\) and \(\rho_{p} < 1\), \cref{eq:osc_transformed} gives \(a_{1}' = 2 \rho_{p}^{J} \cos (J \theta_{p})\) and \(a_{2}' = -\rho_{p}^{2J}\), which imply \(\abs{a_{1}'} < 2\) and \(\abs{a_{2}'} < 1\), so we choose \(I_{a} = 2\) and \(a'\) uses Q2.16.

On the feedforward multiply the Q1.15 input takes the 18-bit port, leaving the 27-bit port for \(b'\).
We restrict the residue to \(\abs{\alpha_{r}} \leq 0.05\).
Over the range presented below, \(\max_{k} \abs{b_{k}'} < 3\)\longonly{ by \cref{eq:osc_ib}}, so \(I_{b} = 3\) and \(b'\) uses Q3.24.

\begin{figure}
  \centering
  \begin{tikzpicture}
    \begin{axis}[
      width=0.74\columnwidth,
      height=4.2cm,
      xmin=5, xmax=150,
      ymin=30, ymax=300,
      xlabel={\(1 / T_{\mathrm{osc}}\) (\unit{\mega\hertz})},
      ylabel={\(\tau\) (\unit{\nano\second})},
      xtick={5, 25, 50, 75, 100, 125, 150},
      xticklabels={\(5\), \(25\), \(50\), \(75\), \(100\), \(125\), \(150\)},
      ytick={30, 100, 200, 300},
      yticklabels={\(30\), \(100\), \(200\), \(300\)},
      tick label style={font=\footnotesize},
      label style={font=\footnotesize},
      ylabel style={rotate=-90},
      colormap/viridis,
      colorbar,
      colorbar style={
        font=\footnotesize,
        title={error (\%)},
        title style={font=\footnotesize, yshift=-2pt},
        ytick={-2.3979, -2.2218, -2, -1.7959, -1.6021, -1.3979},
        yticklabels={\(0.004\), \(0.006\), \(0.01\), \(0.016\), \(0.025\), \(0.04\)},
      },
      point meta min=-2.4306,
      point meta max=-1.3644,
      axis on top,
    ]
      \addplot[
        matrix plot*, point meta={log10(\thisrow{err})},
        mesh/cols=291, mesh/ordering=x varies,
      ] table[x=f, y=tau] {data/osc_error.dat};
      % \closedcycle closes to \(\tau = 0\), which the axis clips back to the floor at 30.
      \addplot[draw=none, fill=white, domain=5:31.831, samples=60]
        {3000 / (pi * x)} \closedcycle;
      \addplot[draw=none, pattern=north east lines, domain=5:31.831, samples=60]
        {3000 / (pi * x)} \closedcycle;
      \fill[white] (61.5,30) rectangle (63.5,300);
      \fill[pattern=north east lines] (61.5,30) rectangle (63.5,300);
      \fill[white] (124,30) rectangle (126,300);
      \fill[pattern=north east lines] (124,30) rectangle (126,300);
      \addplot[white, densely dashed, line width=0.6pt,
               domain=10.61:106.103, samples=60]
        {10000 / (pi * x)};
      \node[anchor=south west, inner sep=1.5pt, font=\footnotesize, text=white]
        at (25,120) {\(Q = 10\)};
    \end{axis}
  \end{tikzpicture}
  \caption{%
    Simulated peak step response error of the damped oscillation correction at the formats of \cref{tab:fixed_point}, for a step of \num{0.5} of full scale, worst case over \(\phi\) at \(\alpha_{r} = \pm 0.05\), on a logarithmic color scale.
    Hatching marks the region excluded by \(Q \geq 3\) and the \SI{1.0}{MHz} guard bands on either side of \(\theta_{p} = m \pi / J\) and \((\theta + \theta_{p}) / 2 = m \pi / J\).
    \(\theta_{p}\) depends on the residue, which is not represented, so the bands are drawn at \(\theta_{p} = \theta\); the exclusion must be tested against the fit's \(\theta_{p}\).
    Outside the hatched region and below the dashed \(Q = 10\) contour, the whole of \(\abs{\alpha_{r}} \leq 0.05\) is correctable at every \(\phi\); above the contour the least favorable \(\phi\) permits \(\abs{\alpha_{r}} \approx 1 / (2 Q)\).
  }\label{fig:osc_error}
\end{figure}
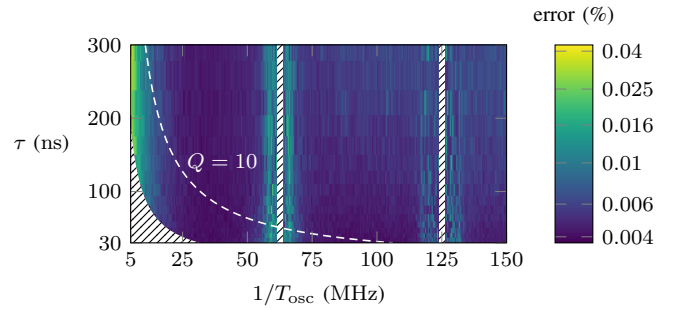

These formats cover \(\SI{5}{MHz} \leq 1/T_{\mathrm{osc}} \leq \SI{150}{MHz}\) at \(Q \geq 3\), \(\SI{30}{ns} \leq \tau \leq \SI{300}{ns}\), and \(\abs{\alpha_{r}} \leq 0.05\) subject to the conditions that keep \(p_{1,2}\) a conjugate pair inside the unit circle\longonly{, \cref{eq:osc_conj_restrict,eq:osc_unit_restrict}} (which hold the correctable \(\abs{\alpha_{r}}\) below \(0.05\) above \(Q = 10\) at unfavorable \(\phi\)), and the \(\min_{\omega} \abs{B'} \geq 2^{-11}\) that Q1.26 sets.
We exclude \SI{1.0}{MHz} on either side of the points where \(\theta_{p}\) and \((\theta + \theta_{p}) / 2\) equal \(m \pi / J\) (\SI{62.5}{MHz} and \SI{125}{MHz}).
A resonance inside a guard band can instead be corrected by a variant of the section with one added feedback register, \(L = 5\) and \(J = 10\), whose bands fall at multiples of \SI{50}{MHz} and are disjoint from those at \(J = 8\), at the cost of five cycles of latency, four feedforward taps, and re-evaluated formats.
\Cref{fig:osc_error} shows the simulated error over the covered region in \(\tau\) and \(T_{\mathrm{osc}}\).
Over that range, the joint \(\tau\) and \(T_{\mathrm{osc}}\) contribution \((2 \abs{\alpha_{r}} / e) (\tau / T_{s}) \mathcal{B}_{z}\) stays below \(0.01\%\), and \(2 \mathcal{E}\) and \(\lVert E \rVert_{\infty}\) are approximately \(t_{\mathrm{step}}\)\longonly{, by the bounds of \cref{sec:pole_sensitivity,sec:zero_sensitivity,sec:param_sensitivity,sec:cancellation}}.

The bias-tee resonances of Guo et al.~\cite[Table~I]{guo2024universal}, at \SIrange{2.6}{3.0}{MHz} with \(\tau\) of \SIrange{66}{84}{ns} and \(Q\) below 1, lie below this range, where proximity to DC raises the bound on \(2 \mathcal{E}\) to about 0.5\%.
We have observed oscillations within the selected band in our own measurements of flux lines.

% \cref{eq:tail_bits} scales \(\tau\) with \(2^{F_{a}}\), and the tail needs \(F_{a} = 16.4\) at \(\tau = \SI{500}{ns}\) and the worst \(\alpha = +0.4\), giving \SI{749}{ns} at 17 bits and \SI{374}{ns} at 16.
% The reduced \cref{eq:tail_bits} overstates the reach by the bits quoted after it; the exact form gives \SI{744}{ns} and \SI{369}{ns}, so the figures below are rounded to two significant figures.
% The biquad as two real tails is not covered by \cref{eq:tail_bits}: two real poles leave \(J \abs{a_{2}'} \abs{p_{1}^{-J} - p_{2}^{-J}}\) in the denominator of \cref{eq:pole_sensitivity}, the degeneracy of \cref{eq:osc_pole_bound} at \(\theta_{p} = m \pi / J\) carried to the real axis, so the pole sensitivity grows without bound as the two time constants approach each other.
The SOS can also compensate a single exponential tail.
Its Q formats have wider range and better precision than those of the FOS except for \(F_{a}\), which is 16 rather than 17; by \cref{eq:tail_bits}, this shortens the \(\tau\) reach at \(t_{\mathrm{step}} = 0.1\%\) from roughly \SIrange{740}{370}{ns}.

% TODO resurrect if we add bounce compensation support
% \subsubsection{Bounce Compensation}

\subsubsection{FIR}

The Q format analysis for the FIR filter differs from that for the IIR filters.
To determine the fractional precision, we consider the coefficient quantization that limits the output error to less than half an output LSB, which is the error contributed by the final convergent round.
For an input at frequency \(\omega\) and amplitude \(X\), the error amplitude is \(X \abs{\sum_{k} \delta b_{k} e^{-j k \omega}} \leq X \sum_{k} \abs{\delta b_{k}} \leq X N_{b} \Delta b / 2\) under round to nearest.
Under full scale at \(X = 1\), the worst-case output error is \(N_{b} \Delta b / 2\).
Therefore,
\begin{equation}\label{eq:fir_bits}
  F_{b} \geq F_{x} + \log_{2} N_{b}.
\end{equation}
For \(F_{x} = 15\) and \(N_{b} = 20\), this gives \(F_{b} = 19.3\).

% Parseval alternative to \cref{eq:fir_bits}, kept for possible later use.
% \(\delta b_{k}\) independent and uniform on \([-\Delta b / 2, \Delta b / 2]\) gives \(\overline{\delta b_{k}^{2}} = \Delta b^{2} / 12\), so the RMS is \(\Delta b \sqrt{N_{b} / 12}\), below \(N_{b} \Delta b / 2\) by \(\sqrt{3 N_{b}}\).
% The peak over \(\omega\) exceeds that RMS by roughly \(\sqrt{2 \ln N_{b}}\), so the requirement lies between the two.
% This worst case requires the \(N_{b}\) coefficient errors to align in phase at one frequency.
% Parseval's theorem equates the mean square of the error response over \(\omega\) to \(\sum_{k} \abs{\delta b_{k}}^{2}\), which for independent errors replaces \(N_{b} \Delta b / 2\) with an RMS of \(\Delta b \sqrt{N_{b} / 12}\) and gives
% \begin{equation}
%   F_{b} \geq F_{x} + \frac{1}{2} \log_{2} \frac{N_{b}}{3},
% \end{equation}
% or 16.4 bits.

% \(H(e^{j\omega}) = \sum_{k} b_{k} e^{-jk\omega}\) is a Fourier series, so \(b_{k} = \left[1/(2 \pi)\right] \int_{-\pi}^{\pi} H(e^{j\omega}) e^{jk\omega} \, d\omega\), and \(\abs{e^{jk\omega}} = 1\) gives the bound.
Each tap is a Fourier coefficient of the frequency response, so
\begin{equation}
  \max_{k} \abs{b_{k}} \leq \frac{1}{2 \pi} \int_{-\pi}^{\pi} \abs{H(e^{j\omega})} \, d\omega,
\end{equation}
and \(I_{b}\) is set by the band-averaged magnitude of the correction.
% TODO \(I_{b} = 3\) is provisional; fixing it requires \(\max_{k} \abs{b_{k}}\) over fits to measured lines.
We take \(I_{b} = 3\), which allows a band-averaged magnitude up to 4 and matches the coefficient range of the HDAWG's FIR~\cite[Table~5.45]{zihdawg}.
Under the worst-case bound on \(F_{b}\), \(b\) uses Q3.20.

\section{Evaluation}\label{sec:eval}

The filters are deployed on QubiC's ZCU216 platform, an XCZU49DR UltraScale+ RFSoC, and built with Vivado 2022.2.
Resource counts are post-implementation with timing met at \(f_{\mathrm{clk}} = \SI{500}{MHz}\).
The Fmax of each module is \SI{775}{MHz}, the maximum clock frequency of both the global clock buffer and the DSP48E2 on the XCZU49DR-2E at \(V_{\mathrm{CCINT}} = \SI{0.85}{V}\)~\cite{ds926}.
To determine Fmax, modules were timed out-of-context with registered ports and a global clock buffer, with the clock applied directly to the port and no MMCM in the path.
An MMCM in the clock path adds about \SI{0.01}{ns} of clock uncertainty at \SI{775}{MHz} and was verified not to change the reported Fmax.

\subsection{Latency}

Our modules achieve the latencies in \cref{tab:eval}, expressed as a cycle count \(\ell\); the corresponding time is \(\ell / f_{\mathrm{clk}}\), and the FIR latency is \(N_{b} + 8\) cycles.
Low latency is important when mid-circuit measurement is used, and latency is thus kept as low as possible without degrading throughput or other critical features, such as rounding and saturation.
Outside that context, the additional filter latency does not degrade QubiC's capabilities.

\begin{table}\centering
  \caption{Filter latency \(\ell\) at QubiC's \(f_{\mathrm{clk}} = \SI{500}{MHz}\) and post-implementation resource usage at \(M = 2\).}\label{tab:eval}
  \begin{tabular}{l r r r r r r}
    \toprule
    & \multicolumn{2}{c}{latency} & \multicolumn{4}{c}{resources}     \\
    \cmidrule(lr){2-3} \cmidrule(lr){4-7}
    filter               & cycles & ns & DSP & logic LUT & SRL & FF   \\
    \midrule
    integrator           & 20     & 40 & 10  & 178       & 24  & 955  \\
    FOS                  & 24     & 48 & 20  & 97        & 56  & 1009 \\
    SOS                  & 33     & 66 & 38  & 166       & 120 & 1834 \\
    \(\mathrm{FIR}[20]\) & 28     & 56 & 40  & 24        & 160 & 1552 \\
    \bottomrule
  \end{tabular}
\end{table}

\subsection{Resource Usage}

Our filters consume primarily DSP48E2 slices, \cref{tab:eval}, with little additional fabric logic.
Moreover, because the datapaths stay largely within DSP columns and their dedicated internal routing, they place correspondingly little demand on general fabric routing.
The general IIR DSP use is
\begin{equation}
  \mathrm{DSP} = M N_{b} + N_{a} L M^{2}.
\end{equation}
For the integrator, however, \(N_{a} M\) can be subtracted from this count, since the feedback tap is unity and does not require a multiplier.
Therefore, the DSP use of the integrator is \(M(2 M + 1)\), that of the FOS \(M(4 M + 2)\), that of the SOS \(M(8 M + 3)\), and that of the FIR \(N_{b} M\).
The quadratic dependency of DSP use on \(M\) in the IIR is the resource cost of the scattered look-ahead transformation.

The cascade of \cref{sec:arch}, comprising an integrator, one SOS, and a 20-tap FIR, therefore uses 88 DSP slices at \(M = 2\).
Instantiating one cascade on each of the XCZU49DR's 16 DAC channels, the most flux lines one controller can drive, uses 1408 DSP slices, one third of the device's 4272~\cite{xmp105}.

\subsection{Hardware Validation}\label{sec:validation}

Our filters were tested in a bit-accurate functional simulation and on hardware to ensure correctness.
They were also used in hardware to correct bias-tee line distortion, \cref{fig:validation}, with only the integrator and FIR active: the integrator time constant, \SI{14}{\us}, was fit to the droop of an uncorrected square step, and the FIR taps were fit to the residual step measured after the integrator.
The captures confirm the filters operate as intended on real distortion; the residual above \(t_{\mathrm{step}}\) reflects our preliminary calibration procedure rather than the filters, whose accuracy is established by the simulations of \cref{sec:fixed_point}, and an improved procedure is left to future work.

% Keysight DSOV084A, 8 vertical divisions, 100003 samples over a 10 us record
% with the trigger 10.001% in, so trigfrac is the same in both panels.
% vscale is rounded from the captured 42.4 mV/div down to 40 mV/div.
% The left panel shows the whole record at the captured 1 us/div; the right
% panel reuses the captures but sets hscale and hpos to zoom in on the rising
% edge, which then sits about 1.4 divisions from the left.
% nth=20 decimates the full-record panel to about 5000 points per trace; the
% zoomed window already holds roughly 1000 points.
\begin{figure*}[t]
  \centering
  \begin{subfigure}{0.48\textwidth}
    \centering
    \scopeplot{
      file={data/yq_cbt12_1g_room_temp_gauss_uncorrected.csv},
      file2={data/yq_cbt12_1g_room_temp_gauss_corrected.csv}, skip=48,
      vscale=4.00e-2, vcenter=6e-2, vdiv=8,
      hscale=1.00e-6, hpos=0, trigfrac=0.10001, nth=20,
      input={DC \qty{50}{\ohm}}, bw=4.00e9, rate=1.000e10,
      width=0.86\linewidth, height=100pt,
    }
    \caption{}\label{fig:validation_full}
  \end{subfigure}
  \hfill
  \begin{subfigure}{0.48\textwidth}
    \centering
    \scopeplot{
      file={data/yq_cbt12_1g_room_temp_gauss_uncorrected.csv},
      file2={data/yq_cbt12_1g_room_temp_gauss_corrected.csv}, skip=48,
      vscale=4.00e-2, vcenter=8.48e-2, vdiv=8,
      hscale=10e-9, hpos=-4e-9, trigfrac=0.10001,
      input={DC \qty{50}{\ohm}}, bw=4.00e9, rate=1.000e10,
      width=0.86\linewidth, height=100pt,
    }
    \caption{}\label{fig:validation_edge}
  \end{subfigure}
  \caption{%
    Oscilloscope captures of an \SI{8}{\us} Gaussian-edged pulse through a \(\mathrm{YQ}\text{-}\mathrm{CBT}12\text{-}1\mathrm{G}\) bias tee at room temperature, uncorrected (dotted) and corrected (solid) by the FPGA filters.
    \subref{fig:validation_full} shows the full record and \subref{fig:validation_edge} the rising edge, with a corrected 10\% to 90\% rise time of \SI{4.2}{ns}.
    The filters were calibrated on a square step and evaluated on the filtered edge.
    Relative to the mean corrected pulse top, the RMS (peak) deviation over \SIrange{3}{30}{ns} after the edge (50\% crossing) is 13.8\% (27\%) uncorrected and 0.31\% (1.7\%) corrected, and over \SI{30}{ns} to \SI{7.9}{\us} it is 26\% (42\%) and 0.09\% (1.7\%).
  }\label{fig:validation}
\end{figure*}

\section{Conclusions and Future Work}\label{sec:conclusion}

We presented an FIR/IIR predistortion cascade for superconducting qubit flux lines, running at \SI{1}{GS/s} on QubiC's \SI{500}{MHz} fabric clock.
Scattered look-ahead pipelining with look-ahead \(J = L M\)~\cite{parhi1989block} keeps every transformed feedback tap within its own polyphase branch, so the recursion maps onto DSP48E2 cascade routing without fabric arithmetic and without altering the target transfer function.
The fixed-point analysis translates the parameter ranges and correction accuracy required by the application into coefficient and accumulator fixed-point formats.
The resulting filters fit on all 16 DAC channels of the ZCU216 and operate as intended on hardware, correcting characteristic bias-tee distortion.
In future work we plan to integrate the filters with cryoscope-based calibration on superconducting qubits and add a dedicated recursive section for reflections whose delay exceeds the reach of the FIR\@.
Our implementation will be made publicly available upon publication.

\iflong % appendix, long version only
\appendices
% IEEEtran's \appendices does not inform cleveref, which would otherwise print "Section A".
\crefalias{section}{appendix}
\crefalias{subsection}{appendix}

\section{Stability}\label{sec:stability}

Unconditional filter stability requires \(\abs{\hat{p}_{i}} < 1\) for all transfer function poles after scattered look-ahead transformation and coefficient quantization.
The output of filters satisfying \(\abs{\hat{p}_{i}} = 1\) can remain bounded for suitable inputs.

Combining \(A(z)\) in \cref{eq:iir_h} with \(\Phi_{J}(z)\) in \cref{eq:phi} yields the transformed denominator of our transfer function
\begin{equation}
  A'(z) = A(z) \Phi_{J}(z) = 1 - \sum_{k=1}^{N_{a}} a'_{k} z^{-k J}.
\end{equation}
% The factored form follows from the fundamental theorem of algebra.
% \(a'_{N_{a}}\) is the coefficient of the factored form because when we expand the factors, it is the coefficient of the \(\zeta^{N_{a}}\) term.
The substitution \(\zeta \coloneq z^{-J}\) makes this a polynomial of degree \(N_{a}\) in \(\zeta\) rather than a sparse one in \(z^{-1}\),
\begin{equation}\label{eq:a_prime_v}
  A'(\zeta) = 1 - \sum_{k=1}^{N_{a}} a'_{k} \zeta^{k}
            = -a'_{N_{a}} \prod_{l=1}^{N_{a}} (\zeta - \zeta_{l}),
\end{equation}
whose roots are \(\zeta_{i} = p_{i}^{-J}\), with \(p_{i}\) the roots of \(A(z)\).

\subsection{High-Pass Droop}\label{sec:high_pass_stability}

\(p_{1} = 1\), so \cref{eq:high_pass_aprime} gives \(a_{1}' = 1\), which is exactly representable by omitting the feedback multiplication, so \(\abs{\hat{p}_{1}} = 1\).
This filter has infinite DC gain and is thus not unconditionally stable.
However, its output is bounded for suitable inputs.
Specifically, if we restrict the quantized coefficients to preserve the structure of \cref{eq:high_pass_bprime},
\begin{subequations}
  \begin{align}
    \hat{b}_{0}' &= 1,\label{eq:stability_high_pass_restrict_b0}\\
    \hat{b}_{1}' &= \hat{b}_{2}' = \cdots = \hat{b}_{J-1}',\label{eq:stability_high_pass_restrict_b1}\\
    1 + \hat{b}_{J}' &= \hat{b}_{k=1, \ldots, J-1}',\label{eq:stability_high_pass_restrict_bJ}
  \end{align}
\end{subequations}
then, with zero initial conditions,
% The transformed difference equation is
% \begin{align}
%   y[n] &= y[n-J] + x[n] + \hat{b}_{1}' \sum_{k=1}^{J-1} x[n-k]
%          + \left(\hat{b}_{1}' - 1\right) x[n-J]\\
%        &= y[n-J] + w[n],
% \end{align}
% with
% \begin{equation}
%   w[n] \coloneq (x[n] - x[n-J]) + \hat{b}_{1}' \sum_{k=1}^{J} x[n-k].
% \end{equation}
% Therefore, recursively substituting in \(y[n-J]\), \(y[n-2J]\), etc. and imposing a zero initial input and output state, we eventually get the result.
\begin{equation}
  y[n] = x[n] + \hat{b}_{1}' \sum_{k=0}^{n-1} x[k],
\end{equation}
so the output is bounded when the accumulated input is bounded.

\subsection{Exponential Tail}

% With \(1 + \alpha > 0\), \(\abs{p_{1}} < 1\) is \(-(1 + \alpha) < \rho + \alpha < 1 + \alpha\).
% The right inequality is \(\rho < 1\), which always holds, and the left rearranges to \cref{eq:exp_tail_alpha_restrict}.
Stability in \cref{eq:tail_pre}, \(\abs{p_{1}} < 1\), further restricts \(\alpha\) to
\begin{equation}\label{eq:exp_tail_alpha_restrict}
  \alpha > -\frac{1}{2} \left(1 + e^{-T_{s}/\tau}\right).
\end{equation}
% \(\hat{A}'(\zeta) = 1 - \hat{a}_{1}' \zeta\) has root \(\zeta = 1 / \hat{a}_{1}'\), so \(\hat{p}_{1}^{J} = \hat{a}_{1}'\) and \(\abs{\hat{p}_{1}} < 1 \iff \abs{\hat{a}_{1}'} < 1\).
% The triangle inequality gives \(\abs{\hat{a}_{1}'} \leq \abs{p_{1}}^{J} + \Delta a' / 2\), so \(\Delta a' < 2 \left(1 - \abs{p_{1}}^{J}\right)\) suffices, and \cref{eq:tail_zeros_poles,eq:integrator_coeff} substitute the physical parameters for \(p_{1}\).
% When the feedback format has \(I \geq 2\), so that \(\abs{\hat{a}_{1}'} \geq 1\) is representable, the condition is also necessary in the worst rounding case.
Quantization preserves stability if
\begin{equation}\label{eq:exp_stability}
  \Delta a'
  < 2 \left(1 -
    \abs*{\frac{e^{-T_{s}/\tau} + \alpha}{1 + \alpha}}^{J}\right).
\end{equation}
% \(\abs{p_{1}} < 1\) keeps the right-hand side of \cref{eq:exp_stability} positive.

\subsection{Damped Oscillation}

% Start with the realized transfer function denominator, \(\hat{A}'(\zeta) = 1 - \hat{a}_{1}' \zeta - \hat{a}_{2}' \zeta^{2}\).
% The reciprocal polynomial is \(P(\eta) = \eta^{2} \hat{A}'(1/\eta) = \eta^{2} - \hat{a}_{1}' \eta - \hat{a}_{2}'\), in \(\eta \coloneq 1 / \zeta\).
% The Jury stability criteria are \(P(1) > 0\), \(P(-1) > 0\), and \(\abs{P(0)} < 1\).
Applying the Jury stability criterion~\cite{jury1962} to the reciprocal polynomial of \cref{eq:a_prime_v}, whose roots are \(p_{i}^{J}\), yields the necessary and sufficient conditions for stability:
\begin{equation}
  \hat{a}_{1}' + \hat{a}_{2}' < 1,
    \quad \hat{a}_{2}' - \hat{a}_{1}' < 1,
    \quad \abs{\hat{a}_{2}'} < 1.
\end{equation}
% Round to nearest gives \(\abs{\delta a_{k}'} \leq \Delta a' / 2\).
% The worst case of each criterion, counting \(\abs{\hat{a}_{2}'} < 1\) as two, is \(\Delta a' < 1 - a_{1}' - a_{2}'\), \(\Delta a' < 1 + a_{1}' - a_{2}'\), \(\Delta a' < 2 (1 - a_{2}')\), and \(\Delta a' < 2 (1 + a_{2}')\).
% The third is the sum of the first two, which are both positive whenever the exact filter is stable, so it never governs; the first two differ only in the sign of \(a_{1}'\) and merge into \(\Delta a' < 1 - a_{2}' - \abs{a_{1}'}\).
% With \(p_{1} p_{2} = c_{2}/c_{0}\) and \(p_{1} + p_{2} = -c_{1}/c_{0}\) from \cref{eq:osc_num}, \(\sigma = (p_{1} p_{2})^{J/2}\) and \(\xi = (p_{1} + p_{2}) / (2 \sqrt{p_{1} p_{2}})\) give \(a_{2}' = -(p_{1} p_{2})^{J} = -\sigma^{2}\) and \(a_{1}' = p_{1}^{J} + p_{2}^{J} = 2 \sigma \mathcal{T}_{J}(\xi)\).
% The bound is sufficient but not necessary, since it assumes each \(\delta a_{k}'\) attains \(\pm \Delta a' / 2\) with the worst sign.
Since \cref{eq:osc_transformed} gives \(a_{1}' = p_{1}^{J} + p_{2}^{J} = 2 \sigma \mathcal{T}_{J}(\xi)\) and \(a_{2}' = -\left(p_{1} p_{2}\right)^{J} = -\sigma^{2}\), the worst case of these criteria under round to nearest shows that quantization preserves stability if
\begin{equation}\label{eq:osc_stability_bound}
  \Delta a' < \min\{
    1 + \sigma^{2} - \abs{2 \sigma \mathcal{T}_{J}(\xi)},\,
    2 (1 - \sigma^{2})
  \},
\end{equation}
where \(\mathcal{T}_{J}(\xi) = \cos \left[J \arccos(\xi)\right]\) is the Chebyshev polynomial of the first kind and
\begin{subequations}\label{eq:osc_sigma_xi}
\begin{align}
  \sigma &\coloneq \left(p_{1} p_{2}\right)^{J/2}
          = \left(\frac{\rho^{2} + 2 \alpha_{r} \rho \cos(\theta - \phi)}
                       {1 + 2 \alpha_{r} \cos\phi}\right)^{J/2},\\
  \begin{split}
    \xi &\coloneq \frac{p_{1} + p_{2}}{2 \sqrt{p_{1} p_{2}}}\\
        &= \frac{\rho \cos\theta
                  + \alpha_{r} \cos\phi + \alpha_{r} \rho \cos(\theta - \phi)}
                 {\sqrt{\left(1 + 2 \alpha_{r} \cos\phi\right)
                        \left(\rho^{2}
                              + 2 \alpha_{r} \rho \cos(\theta - \phi)\right)}}.
  \end{split}
\end{align}
\end{subequations}

% \cref{eq:osc_num} makes the discriminant \(c_{1}^{2} - 4 c_{0} c_{2} = 4 \left[\left(\operatorname{Re} \beta\right)^{2} - 2 \rho \sin\theta \operatorname{Im} \beta - \rho^{2} \sin^{2}\theta\right]\), which \(\left(\operatorname{Re} \beta\right)^{2} = \abs{\beta}^{2} - \left(\operatorname{Im} \beta\right)^{2}\) collapses to \(\abs{\beta}^{2} - \left(\operatorname{Im} \beta + \rho \sin\theta\right)^{2}\).
% A negative discriminant is therefore \(\abs{\beta} < \abs{\operatorname{Im} \beta + \rho \sin\theta}\), whose right side is positive because it exceeds \(\abs{\beta} \geq \abs{\operatorname{Im} \beta}\).
% \cref{eq:osc_num} also gives \(1 - p_{1} p_{2} = \left[\left(1 - \rho^{2}\right) + 2 \operatorname{Re} \beta\right] / c_{0}\).
% \(\operatorname{Im} \beta\) and \(\operatorname{Re} \beta\) are both at least \(-\abs{\beta} = -\abs{\alpha_{r}} \abs{1 - \rho e^{j \theta}}\), which gives the \(\phi\)-free \(\abs{\alpha_{r}} < \rho \sin\theta / \left(2 \abs{1 - \rho e^{j \theta}}\right)\) and \(\abs{\alpha_{r}} < \left(1 - \rho^{2}\right) / \left(2 \abs{1 - \rho e^{j \theta}}\right)\).
\Cref{eq:sos} takes \(p_{1,2}\) from a quadratic with real coefficients, so the two are either a conjugate pair or both real.
At \(\alpha_{r} = 0\) they coincide with the zeros \(q_{1,2}\), which lie off the real axis, and they turn real only once the residue is large enough to bring them together, at which point the section no longer describes a decaying oscillation.
We therefore restrict the analysis to the conjugate-pair branch, which, with
\begin{equation}\label{eq:osc_beta}
  \beta \coloneq \alpha_{r} e^{j \phi} \left(1 - \rho e^{-j \theta}\right),
\end{equation}
holds exactly when
\begin{equation}\label{eq:osc_conj_restrict}
  \abs{\beta} - \operatorname{Im} \beta < \rho \sin\theta.
\end{equation}
\Cref{eq:osc_num} makes the discriminant \(c_{1}^{2} - 4 c_{0} c_{2} = 4 \left[\abs{\beta}^{2} - \left(\operatorname{Im} \beta + \rho \sin\theta\right)^{2}\right]\), which is negative under \cref{eq:osc_conj_restrict}.
% \(g[0] = 1 + 2 \alpha_{r} \cos\phi = c_{0}\), so \(2 \abs{\alpha_{r}} < 1\) keeps it positive.
\(c_{0}\) is the initial value \(g[0]\) of the step response, positive whenever the ringing amplitude \(2 \abs{\alpha_{r}}\) is less than the unit step, so the gain \(\kappa = 1 / c_{0}\) of \cref{eq:sos} is positive.
A conjugate pair has \(\abs{p_{1}} = \abs{p_{2}} = \sqrt{p_{1} p_{2}}\), and \cref{eq:osc_num} gives \(c_{0} \left(1 - p_{1} p_{2}\right) = \left(1 - \rho^{2}\right) + 2 \operatorname{Re} \beta\), so \(H\) is stable when
\begin{equation}\label{eq:osc_unit_restrict}
  2 \operatorname{Re} \beta > -\left(1 - \rho^{2}\right).
\end{equation}
\Cref{eq:osc_conj_restrict} additionally ensures that \(\sigma\) and \(\xi\) are real with \(\abs{\xi} < 1\), and \cref{eq:osc_unit_restrict} makes \(\sigma < 1\).

% The first term is the smaller of the two when \(1 + \sigma^{2} - 2 \sigma \abs{\mathcal{T}_{J}(\xi)} < 2 (1 - \sigma^{2})\), which rearranges to the threshold below.
% \(\xi = \cos(m \pi / J)\) is \(J \arccos \xi = m \pi\), so the phases of \(p_{1,2}\) advance by a multiple of \(\pi\) over the \(J\)-sample delay and their \(J\)th powers agree.
% Near a center, \(1 - \abs{\mathcal{T}_{J}(\xi)} \approx J^{2} \left(\theta_{p} - m \pi / J\right)^{2} / 2\), so the first term is \((1 - \sigma)^{2} + \sigma J^{2} \left(\theta_{p} - m \pi / J\right)^{2}\).
Define \(\rho_{p}\) and \(\theta_{p}\) by
\begin{equation}\label{eq:osc_pole_polar}
  p_{1,2} = \rho_{p} e^{\pm j \theta_{p}},
\end{equation}
so that \(\theta_{p} = \arccos(\xi)\), \(\sigma = \rho_{p}^{J}\), and \(\mathcal{T}_{J}(\xi) = \cos(J \theta_{p})\).
The second term of \cref{eq:osc_stability_bound} is the smaller of the two unless \(\abs{\mathcal{T}_{J}(\xi)} > (3 \sigma^{2} - 1) / (2 \sigma)\), which confines the first term to notches centered on \(\theta_{p} = m \pi / J\) for integer \(m\).
At a notch center the feedback delay \(J T_{s}\) spans an integer number of half periods of \(p_{1,2}\), the correction's own oscillation, so \(p_{1}^{J} = p_{2}^{J}\).

\section{Pole Sensitivity}\label{sec:pole_sensitivity}

In \cref{eq:a_prime_v}, perturb each coefficient \(a'_{k}\) by \(\delta a'_{k}\) and require \(\hat{A}'\) to vanish at the displaced root, as in the root sensitivity analysis of Kaiser~\cite{kaiser1966digital}, \cite[Sec.~6.8.1]{oppenheim2010}.
Taking \(\partial A' / \partial \zeta\) from the product form of \cref{eq:a_prime_v} and \(\partial A' / \partial a'_{k} = -\zeta^{k}\) from the sum, then converting \(\delta \zeta_{i}\) to \(\delta p_{i}\) through \(\delta \zeta_{i} / \zeta_{i} = -J\, \delta p_{i} / p_{i}\), gives
\begin{equation}\label{eq:pole_sensitivity}
  \frac{\delta p_{i}}{p_{i}}
    = \frac{\sum_{k=1}^{N_{a}} p_{i}^{-(k-1) J}\, \delta a'_{k}}
           {J a'_{N_{a}} \prod_{l \neq i} \left(p_{i}^{-J} - p_{l}^{-J}\right)}.
\end{equation}

\subsection{High-Pass Droop}

As stated in \cref{sec:high_pass_stability}, \(\hat{p}_{1} = p_{1} = 1\), so the pole location is unaffected by transformed coefficient quantization.

\subsection{Exponential Tail}

For an exponential tail, \(N_{a}=1\), \(a'_{1} = p_{1}^{J}\), so
\begin{equation}\label{eq:exp_pole_sensitivity}
  \frac{\delta p_{1}}{p_{1}} = \frac{\delta a'_{1}}{J p_{1}^{J}}.
\end{equation}
Imposing round to nearest, we get
\begin{equation}\label{eq:exp_pole_bound}
  \abs*{\frac{\delta p_{1}}{p_{1}}} \leq \frac{\Delta a'}{2 J \abs{p_{1}}^{J}}.
\end{equation}

\subsection{Damped Oscillation}

For a damped oscillation, \(N_{a} = 2\), and \cref{eq:osc_transformed} gives \(a'_{2} = -(p_{1} p_{2})^{J}\), which collapses the denominator of \cref{eq:pole_sensitivity} to \(J (p_{i}^{J} - p_{l}^{J})\) for \(l \neq i\).
\Cref{eq:osc_conj_restrict} makes the poles a conjugate pair in the form of \cref{eq:osc_pole_polar}, so, with \(\sigma = \rho_{p}^{J}\),
\begin{equation}\label{eq:osc_pole_sensitivity}
  \frac{\delta p_{1}}{p_{1}}
    = \frac{\delta a_{1}' + \sigma^{-1} e^{-j J \theta_{p}} \delta a_{2}'}
           {2 j J \sigma \sin (J \theta_{p})},
\end{equation}
and \(\delta p_{2} / p_{2} = \overline{\delta p_{1} / p_{1}}\), since the \(a_{k}'\) are real.
At \(\theta_{p} = m \pi / J\), \(\sin (J \theta_{p}) = 0\) and \cref{eq:osc_pole_sensitivity} does not impose a bound.
\Cref{eq:osc_pole_sensitivity} also fails to impose a valid bound in the neighborhood of these points, since the two roots of \cref{eq:a_prime_v} converge and a perturbation comparable to their separation displaces them by more than the first-order estimate.
% With \(\mathcal{D} = -d\) and \(\delta \mathcal{D} = d\), the roots move by \(\sqrt{d}\) against a first-order estimate of \(\sqrt{d} / 2\); closer to the degeneracy the first-order estimate instead overstates the displacement without limit.
We therefore bound the displacement of the roots of \cref{eq:a_prime_v} directly, which holds for every \(\theta_{p}\).
% \cref{eq:osc_pole_sensitivity} is first order in \(\delta a_{k}'\), so it fails where its denominator vanishes; bounding the roots of \cref{eq:a_prime_v} directly avoids that and does not require a separate degenerate case.
% \(a_{1}' = 2 \sigma \mathcal{T}_{J}(\xi)\) and \(a_{2}' = -\sigma^{2}\) follow from \cref{eq:osc_transformed,eq:osc_pole_polar}, so \(\mathcal{D} = 4 \sigma^{2} \left[\mathcal{T}_{J}(\xi)^{2} - 1\right]\), which vanishes at \(\theta_{p} = m \pi / J\).
% \(\mathcal{D}\) is real and non-positive, so the square-root difference is \(\bigl|\sqrt{\abs{\mathcal{D} + \delta \mathcal{D}}} - \sqrt{\abs{\mathcal{D}}}\bigr|\) where \(\mathcal{D} + \delta \mathcal{D} \leq 0\) and \(\sqrt{\delta \mathcal{D}}\) elsewhere.
% Both are at most \(\sqrt{\abs{\delta \mathcal{D}}}\); the first is also \(\abs{\delta \mathcal{D}} / \left(\sqrt{\abs{\mathcal{D} + \delta \mathcal{D}}} + \sqrt{\abs{\mathcal{D}}}\right) \leq \abs{\delta \mathcal{D}} / \sqrt{\abs{\mathcal{D}}}\), and the second satisfies that too since \(\delta \mathcal{D} \geq \abs{\mathcal{D}}\) there.
% Dividing by \(J \abs{p_{i}^{J}} = J \sigma\) gives the result; that division is the only step taken to first order, and near the degeneracy it understates \(\abs{\delta p_{i} / p_{i}}\) by a relative amount of roughly \((1 - 1/J) \sqrt{\mathcal{G}} / (2 \sigma)\), which is of order \(\sqrt{\Delta a'}\).
% Dropping \(\Delta a' / (4 J \sigma)\), or halving the second branch as a first-order treatment would, each break the bound at first order.
The reciprocal polynomial of \cref{eq:a_prime_v} has roots \(p_{i}^{J} = \left(a_{1}' \pm \sqrt{\mathcal{D}}\right) / 2\), with discriminant \(\mathcal{D} \coloneq a_{1}^{\prime\,2} + 4 a_{2}' = -4 \sigma^{2} \sin^{2}(J \theta_{p})\), so quantization displaces \(p_{i}^{J}\) by \(\delta a_{1}' / 2 \pm \left(\sqrt{\mathcal{D} + \delta \mathcal{D}} - \sqrt{\mathcal{D}}\right) / 2\), where \(\delta \mathcal{D} = \left(\hat{a}_{1}' + a_{1}'\right) \delta a_{1}' + 4\, \delta a_{2}'\).
\(\abs{a_{1}'} \leq 2 \sigma\) and \(\abs{\delta a_{k}'} \leq \Delta a' / 2\) give \(\abs{\delta \mathcal{D}} \leq \mathcal{G}\), and \(\bigl|\sqrt{\mathcal{D} + \delta \mathcal{D}} - \sqrt{\mathcal{D}}\bigr| \leq \min\left(\sqrt{\mathcal{G}},\, \mathcal{G} / \sqrt{\abs{\mathcal{D}}}\right)\), so
\begin{equation}\label{eq:osc_pole_bound}
  \begin{split}
    \abs*{\frac{\delta p_{i}}{p_{i}}}
      &\leq \frac{\Delta a'}{4 J \sigma}
         + \min \left(W,\,
                      \frac{\mathcal{G}}
                           {4 J \sigma^{2} \abs*{\sin (J \theta_{p})}}\right),\\
    W &\coloneq \frac{\sqrt{\mathcal{G}}}{2 J \sigma},
    \qquad
    \mathcal{G} \coloneq \left(2 + 2 \sigma + \frac{\Delta a'}{4}\right) \Delta a',
  \end{split}
\end{equation}
for either pole and every \(\theta_{p}\).
% \cref{eq:osc_num} gives \(1 - p_{1} p_{2} = \left[(1 - \rho^{2}) + 2 \operatorname{Re} \beta\right] / c_{0}\), with \(\operatorname{Re} \beta \leq \abs{\beta}\) and \(c_{0} \geq 1 - 2 \abs{\alpha_{r}}\).
% \(1 - \sigma \approx J (1 - \rho_{p})\), which the same bound puts at roughly \(J T_{s} / \tau = L / (f_{\mathrm{clk}} \tau)\) for a small residue, independent of \(M\).
% \(\abs{\sin (J \theta_{p})} \approx J \abs{\theta_{p} - m \pi / J}\) near a notch, which sets the crossover at \(W\) to within a factor of two.
The branches cross at approximately \(W\) from \(\theta_{p} = m \pi / J\) for integer \(m\), where the two roots of \cref{eq:a_prime_v} converge: inside, they separate as \(\sqrt{\delta \mathcal{D}}\) and the \(W\) term applies; outside, they separate linearly in \(\delta a_{k}'\) and \cref{eq:osc_pole_sensitivity} describes them.
% The second branch reduces to the exponential tail's \(\Delta a' / (2 J)\) only where \(1 - \rho_{p}^{2}\), which \cref{eq:osc_beta,eq:osc_unit_restrict} put below \(\left[\left(1 - \rho^{2}\right) + 2 \abs{\beta}\right] / \left(1 - 2 \abs{\alpha_{r}}\right)\), is small against \(1/J\), so that comparison is not made in the text.

% TODO resurrect if we add bounce compensation support
% \subsection{Bounce Compensation}?

\section{Zero Sensitivity}\label{sec:zero_sensitivity}

The transformed numerator is
\begin{equation}
  B'(z) = B(z) \Phi_{J}(z) = \sum_{k=0}^{N_{b}'-1} b'_{k} z^{-k},
\end{equation}
with \(N_{b}'\) from \cref{eq:nb_prime}.
% Perturbing each \(b'_{k}\) by \(\delta b'_{k}\) and requiring \(\hat{B}'\) to vanish at the displaced root gives \(0 = \left. (\partial B' / \partial w) \right|_{w_{i}} \delta w_{i} + \sum_{k} \left. (\partial B' / \partial b'_{k}) \right|_{w_{i}} \delta b'_{k}\), with \(w_{i} = q_{i}^{-1}\).
% \(B'(w) = \sum_{k} b'_{k} w^{k}\), so the second partial is \(w_{i}^{k} = q_{i}^{-k}\).
% For the first, \(B'(w) = B(w) \Phi_{J}(w)\) gives \(\partial B' / \partial w = \Phi_{J} (\partial B / \partial w) + B (\partial \Phi_{J} / \partial w)\), whose second term vanishes at \(w_{i}\) because \(B(w_{i}) = 0\).
% Hence \(\delta w_{i} = -\sum_{k} q_{i}^{-k} \delta b'_{k} / \left[\Phi_{J}(q_{i}) \left. (\partial B / \partial w) \right|_{w_{i}}\right]\), and \(w_{i} = 1 / q_{i}\) implies \(\delta q_{i} / q_{i} = -\delta w_{i} / w_{i} = -q_{i} \delta w_{i}\).
With \(w \coloneq z^{-1}\), perturbing each \(b'_{k}\) by \(\delta b'_{k}\) and requiring \(\hat{B}'\) to vanish to first order at the displaced root \(w = q_{i}^{-1}\) gives
\begin{equation}\label{eq:zero_sensitivity}
  \frac{\delta q_{i}}{q_{i}}
    = \frac{q_{i} \sum_{k=0}^{N_{b}'-1} q_{i}^{-k}\, \delta b'_{k}}
           {\Phi_{J}(q_{i}) \, \left. \frac{\partial B}{\partial w} \right|_{w = q_{i}^{-1}}}.
\end{equation}
% \(A'(z) = \prod_{l} (1 - p_{l}^{J} z^{-J})\) and \(A(z) = \prod_{l} (1 - p_{l} z^{-1})\), so \(\Phi_{J} = A' / A\) gives \cref{eq:phi_at_zero}.
Since \(A'(z) = A(z) \Phi_{J}(z)\) and \(A'\) factors as \(\prod_{l} (1 - p_{l}^{J} z^{-J})\),
\begin{equation}\label{eq:phi_at_zero}
  \Phi_{J}(q_{i}) = \prod_{l=1}^{N_{a}} \frac{1 - (p_{l}/q_{i})^{J}}{1 - p_{l}/q_{i}}
                  = \prod_{l=1}^{N_{a}} \sum_{k=0}^{J-1} \left(\frac{p_{l}}{q_{i}}\right)^{k}.
\end{equation}

\subsection{High-Pass Droop}

For the high-pass compensation filter, \(N_{b} = 2\), \(N_{a} = 1\), \(N_{b}' = J + 1\), \(p_{1} = 1\), and \(B(w) = 1 - q_{1} w \implies \partial B / \partial w = -q_{1}\), so
\begin{equation}\label{eq:integrator_zero_sensitivity}
  \frac{\delta q_{1}}{q_{1}} = -\frac{\sum_{k=0}^{J} q_{1}^{-k} \delta b_{k}'}{\sum_{k=0}^{J-1} q_{1}^{-k}}.
\end{equation}
% Dropping \(k=0\) and taking the remaining \(\delta b_{k}'\) to a common value leaves \(\sum_{k=1}^{J} q_{1}^{-k} = q_{1}^{-1} \sum_{k=0}^{J-1} q_{1}^{-k}\) in the numerator, which cancels the denominator and leaves \(\delta q_{1} / q_{1} = -\delta b_{1}' / q_{1}\).
\Cref{eq:stability_high_pass_restrict_b0,eq:stability_high_pass_restrict_b1,eq:stability_high_pass_restrict_bJ} make \(\hat{b}_{0}' = 1\) exact, so
\begin{equation}\label{eq:high_pass_restrict_perturbation}
  \delta b_{0}' = 0, \quad \delta b_{1}' = \delta b_{2}' = \cdots = \delta b_{J}',
\end{equation}
and the root sensitivity simplifies to
\begin{equation}
  \delta q_{1} = -\delta b_{1}'.
\end{equation}
Imposing round to nearest yields
\begin{equation}\label{eq:integrator_zero_bound}
  \abs*{\frac{\delta q_{1}}{q_{1}}} \leq \frac{\Delta b'}{2 q_{1}},
\end{equation}
independent of \(J\).

\subsection{Exponential Tail}

With \(B(w) = \kappa (1 - q_{1} w)\),
\begin{equation}\label{eq:tail_zero_sensitivity}
  \frac{\delta q_{1}}{q_{1}}
    = -\frac{\sum_{k=0}^{J} q_{1}^{-k}\, \delta b'_{k}}
            {\kappa \sum_{k=0}^{J-1} (p_{1}/q_{1})^{k}}.
\end{equation}
Round to nearest gives
\begin{equation}\label{eq:tail_zero_bound}
  \abs*{\frac{\delta q_{1}}{q_{1}}}
    \leq \frac{\Delta b'}{2 \kappa}
         \frac{\sum_{k=0}^{J} q_{1}^{-k}}
              {\abs*{\sum_{k=0}^{J-1} (p_{1}/q_{1})^{k}}}.
\end{equation}
% \cref{eq:tail_zeros_poles} gives \(1 - p_{1} = \kappa (1 - q_{1})\) and \(1 - p_{1}/q_{1} = -\alpha \kappa (1 - q_{1}) / q_{1}\), both governed by \(\kappa (1 - q_{1}) = (1 - q_{1}) / (1 + \alpha) \approx T_{s} / \left[(1 + \alpha) \tau\right]\).
% The sums then approach \(J + 1\) and \(J\).
% Equivalently, the pole time constant \(\tau / \kappa = (1 + \alpha) \tau\) spans many samples, which fails only as \(\alpha \to -1\), the limit \cref{eq:exp_tail_alpha_restrict} bounds.
This approaches \(\left(J+1\right) \Delta b' / (2 \kappa J)\) for \(J T_{s} \ll (1 + \alpha) \tau\).

\subsection{Damped Oscillation}

% \(B(w) = \kappa (1 - q_{1} w)(1 - q_{2} w)\), whose \(w\) derivative loses its second term at \(w = q_{1}^{-1}\), leaving \(-\kappa q_{1} (1 - q_{2}/q_{1}) = -\kappa (q_{1} - q_{2})\).
With \(B(w) = \kappa (1 - q_{1} w)(1 - q_{2} w)\) from \cref{eq:sos} and \(N_{b}' = 2 J + 1\),
\begin{equation}\label{eq:osc_zero_sensitivity}
  \frac{\delta q_{1}}{q_{1}}
    = -\frac{q_{1} \sum_{k=0}^{2J} q_{1}^{-k}\, \delta b'_{k}}
            {\kappa \left(q_{1} - q_{2}\right) \Phi_{J}(q_{1})},
\end{equation}
with \(\Phi_{J}(q_{1})\) from \cref{eq:phi_at_zero} and \(\delta q_{2} = \overline{\delta q_{1}}\).
% \(\abs{q_{1}} = \rho\) and \(\abs{q_{1} - q_{2}} = 2 \rho \abs{\sin\theta}\), and the \(\rho\) cancels.
Round to nearest gives
\begin{equation}\label{eq:osc_zero_bound}
  \abs*{\frac{\delta q_{1}}{q_{1}}}
    \leq \frac{\Delta b' \sum_{k=0}^{2J} \rho^{-k}}
              {4 \kappa \abs{\sin\theta} \abs{\Phi_{J}(q_{1})}}.
\end{equation}
% The \(p_{1,2} \to q_{1,2}\), \(\rho \to 1\) limit of \cref{eq:phi_at_zero} reduces this to \(\Delta b' (2J+1) / \left[4 \kappa J \abs{\sin (J \theta)}\right]\), but that moves the notch below from \((\theta + \theta_{p}) / 2 = m \pi / J\) to \(\theta = m \pi / J\), and \cref{eq:osc_sigma_xi} places \(\theta_{p}\) exactly, so the reduction is not used.
Each factor of \cref{eq:phi_at_zero} is a geometric sum that vanishes only where \(p_{l} / q_{1}\) is a nontrivial \(J\)th root of unity; for \(p_{2} / q_{1} = (\rho_{p} / \rho)\, e^{-j (\theta_{p} + \theta)}\) this requires \(\rho_{p} = \rho\) and \((\theta + \theta_{p}) / 2 = m \pi / J\) for integer \(m\), so \(\Phi_{J}(q_{1})\) is smallest near those points.
% \(\Phi_{J}(q_{1}) = 0\) places \(q_{1}\) on an added zero, giving \(B'\) a double root that quantization splits as \(\sqrt{\Delta b'}\), the counterpart of the degeneracy of \cref{eq:osc_pole_sensitivity} at \(\theta_{p} = m \pi / J\).
% The square root branch of \cref{eq:osc_pole_bound} has no counterpart here because \(\sum_{k} (p_{l} / q_{1})^{k}\) vanishes only when \(p_{l} / q_{1}\) is a nontrivial \(J\)th root of unity, so \(\Phi_{J}(q_{1}) = 0\) also requires \(\rho_{p} = \rho\).
% The condition above alone leaves \(\abs{\Phi_{J}(q_{1})} \approx J^{2} \abs{\rho_{p} / \rho - 1} / \left(2 \abs{\sin (m \pi / J)}\right)\), so \cref{eq:osc_zero_bound} stays finite and a guard band on \(\theta\) excludes the degeneracy.
% Since \(B'\) has degree \(2 J\), a cap would come from \(\partial^{2} B' / \partial w^{2} = 2 \left(\partial B / \partial w\right) \left(\partial \Phi_{J} / \partial w\right)\) at the coincidence rather than from a discriminant.

\section{Physical Parameter Sensitivity}\label{sec:param_sensitivity}

Let \(\mathcal{B}_{p}\) and \(\mathcal{B}_{z}\) denote the right-hand sides of the pole and zero bounds, on \(\abs{\delta p_{i} / p_{i}}\) and \(\abs{\delta q_{i} / q_{i}}\), of \cref{sec:pole_sensitivity,sec:zero_sensitivity} for the filter at hand.
% \(\hat{\tau} / \tau = \ln \rho / \ln \hat{\rho}\) and \(\ln \hat{\rho} = \ln \rho + \ln (1 + \delta \rho / \rho)\), so \(\hat{\tau} / \tau = 1 / (1 - \varepsilon)\) with \(\varepsilon\) as defined, giving \(\delta \tau / \tau = \varepsilon / (1 - \varepsilon)\).
% \(\varepsilon / (1 - \varepsilon)\) increases in \(\varepsilon\), so \(\delta \rho / \rho = +\mathcal{B}_{z}\) gives the bound; the opposite excursion gives \(\abs{\varepsilon} / (1 + \abs{\varepsilon})\), which is the smaller of the two for \(\tau \geq 0.72 T_{s}\), that is for \(\rho \geq 0.25\).
% \(\varepsilon_{z} = 1\) is \(\ln (1 + \mathcal{B}_{z}) = -\ln \rho\), which is \(\rho (1 + \mathcal{B}_{z}) = 1\).
% \(\delta \rho / \rho\) is \(\delta q_{1} / q_{1}\) for a real zero and its real part for a complex one, so \(\mathcal{B}_{z}\) bounds it either way.
% To first order \(\varepsilon \to (\tau / T_{s}) (\delta \rho / \rho)\) and \(\varepsilon_{z} \to (\tau / T_{s}) \mathcal{B}_{z}\).
Every filter carries its time constant in the modulus of a zero, \(\abs{q_{1}} = \rho\), and from \cref{eq:integrator_coeff}, \(\ln \rho = -T_{s} / \tau\), so
\begin{equation}\label{eq:tau_from_zero}
  \frac{\delta \tau}{\tau} = \frac{\varepsilon}{1 - \varepsilon},
  \qquad
  \abs*{\frac{\delta \tau}{\tau}} \leq \frac{\varepsilon_{z}}{1 - \varepsilon_{z}},
\end{equation}
where
\begin{equation}\label{eq:tau_eps}
  \varepsilon \coloneq \frac{\tau}{T_{s}} \ln \left(1 + \frac{\delta \rho}{\rho}\right),
  \qquad
  \varepsilon_{z} \coloneq \frac{\tau}{T_{s}} \ln \left(1 + \mathcal{B}_{z}\right).
\end{equation}
\(\varepsilon_{z} = 1\) corresponds to \(\hat{\rho} = 1\), where the zero has been quantized onto the unit circle and the realized filter no longer decays.

\subsection{High-Pass Droop}

\(\tau\) is the only parameter, and \cref{eq:integrator_zero_bound} makes \(\mathcal{B}_{z} = \Delta b' / (2 \rho)\), so \cref{eq:tau_from_zero,eq:tau_eps} give
\begin{equation}\label{eq:integrator_tau_bound}
  \abs*{\frac{\delta \tau}{\tau}} \leq \frac{\varepsilon_{z}}{1 - \varepsilon_{z}},
  \qquad
  \varepsilon_{z} = \frac{\tau}{T_{s}} \ln \left(1 + \frac{\Delta b'}{2 \rho}\right),
\end{equation}
independent of \(J\).

\subsection{Exponential Tail}

The parameters are \(\tau\), which \cref{eq:tau_from_zero} covers with \(\mathcal{B}_{z}\) from \cref{eq:tail_zero_bound}, and \(\alpha\).
% \(\alpha = (q_{1} - p_{1}) / (p_{1} - 1)\) has \(\partial \alpha / \partial q_{1} = 1 / (p_{1} - 1)\) and \(\partial \alpha / \partial p_{1} = (1 - q_{1}) / (p_{1} - 1)^{2}\), and \(1 - p_{1} = \kappa (1 - q_{1})\) clears the square.
Inverting \cref{eq:tail_zeros_poles} for \(\alpha\), the total differential is
\begin{equation}\label{eq:tail_param_sensitivity}
  \delta \alpha
    = \frac{\delta p_{1} / \kappa - \delta q_{1}}{\kappa \left(1 - q_{1}\right)},
\end{equation}
so, with \(\abs{\delta p_{1}} \leq \abs{p_{1}} \mathcal{B}_{p}\) and \(\abs{\delta q_{1}} \leq q_{1} \mathcal{B}_{z}\),
\begin{equation}\label{eq:tail_alpha_bound}
  \abs{\delta \alpha}
    \leq \frac{\abs{p_{1}} \mathcal{B}_{p} / \kappa + q_{1} \mathcal{B}_{z}}
              {\kappa \left(1 - q_{1}\right)}.
\end{equation}
% \cref{eq:exp_pole_bound} makes \(\abs{p_{1}} \mathcal{B}_{p} = \Delta a' / (2 J \abs{p_{1}}^{J-1})\) and \cref{eq:tail_zero_bound} makes \(q_{1} \mathcal{B}_{z} = q_{1} \Delta b' S / (2 \kappa)\), with \(S\) the ratio of its two sums.
% \(1 - p_{1} = \kappa (1 - q_{1})\) sends \(\abs{p_{1}}^{J-1} \to 1\) and \(S \to (J+1)/J\) once \(J (1 - q_{1}) \ll \kappa^{-1}\), and \(1 - q_{1} \to T_{s} / \tau\).
% \(S\) is what the reduced form of \cref{eq:tail_zero_bound} discards; it exceeds \((J+1)/J\) by about \(J T_{s} / \tau\), nearly independent of \(J\), so the limit below is not itself a bound.
With \cref{eq:exp_pole_bound,eq:tail_zero_bound}, for \(J T_{s} \ll (1 + \alpha) \tau\) this approaches \(\tau \left[\Delta a' + (J+1) \Delta b'\right] / \left(2 \kappa^{2} J T_{s}\right)\).

\subsection{Damped Oscillation}

The parameters are \(\tau\), \(T_{\mathrm{osc}}\), \(\alpha_{r}\), and \(\phi\), with \(\mathcal{B}_{p}\) from \cref{eq:osc_pole_bound} and \(\mathcal{B}_{z}\) from \cref{eq:osc_zero_bound}.
\Cref{eq:tau_from_zero} bounds \(\abs{\delta\tau/\tau}\).
% \(T_{\mathrm{osc}} \propto 1/\theta\) gives \(\hat{T}_{\mathrm{osc}} / T_{\mathrm{osc}} = 1 / (1 + \delta \theta / \theta)\) and so the exact form below, whose modulus is largest at \(\delta \theta / \theta = -\varepsilon_{\theta}\).
% Both components are bounded by the modulus of \cref{eq:osc_zero_sensitivity}, and \(\theta = 2 \pi T_{s} / T_{\mathrm{osc}}\) converts \(1/\theta\) to \(T_{\mathrm{osc}} / (2 \pi T_{s})\).
Separating \cref{eq:osc_zero_sensitivity} into \(\delta q_{1}/q_{1} = \delta \rho/\rho + j \delta \theta\) and taking its imaginary part gives
\begin{equation}\label{eq:osc_tau_tosc}
  \frac{\delta T_{\mathrm{osc}}}{T_{\mathrm{osc}}}
    = \frac{-\delta \theta / \theta}{1 + \delta \theta / \theta},
  \qquad
  \abs*{\frac{\delta T_{\mathrm{osc}}}{T_{\mathrm{osc}}}}
    \leq \frac{\varepsilon_{\theta}}{1 - \varepsilon_{\theta}},
\end{equation}
with \(\varepsilon_{\theta} \coloneq T_{\mathrm{osc}} \mathcal{B}_{z} / (2 \pi T_{s})\), since \(\mathcal{B}_{z}\) bounds both \(\abs{\delta \theta}\) and \(\abs{\delta \rho / \rho}\).

The remaining two parameter bounds follow from the residues of \cref{eq:osc_num}, which enter only through the in-phase and quadrature components
\begin{equation}\label{eq:uv_def}
  u \coloneq \alpha_{r} \cos\phi, \qquad v \coloneq \alpha_{r} \sin\phi,
\end{equation}
so that the coefficients \(c_{k}\) are affine in \(u\) and \(v\) and the map inverts in closed form:
% \cref{eq:osc_num} gives \(e_{1} \coloneq p_{1} + p_{2} = -c_{1}/c_{0}\) and \(e_{2} \coloneq p_{1} p_{2} = c_{2}/c_{0}\).
% The identity \(\alpha_{r} \cos(\theta - \phi) = u \cos\theta + v \sin\theta\) makes \(c_{0} = 1 + 2 u\), \(c_{1} = -2 \left[\rho \cos\theta + \rho (u \cos\theta + v \sin\theta) + u\right]\), and \(c_{2} = \rho^{2} + 2 \rho (u \cos\theta + v \sin\theta)\).
% Clearing \(c_{0}\) from \(e_{1}\) and \(e_{2}\) leaves \(2 u (e_{2} - \rho \cos\theta) - 2 \rho v \sin\theta = \rho^{2} - e_{2}\) and \(2 u (e_{1} - \rho \cos\theta - 1) - 2 \rho v \sin\theta = 2 \rho \cos\theta - e_{1}\).
% Subtracting the first from the second eliminates \(v\), giving \(2 u (e_{1} - e_{2} - 1) = A(1) - D(1)\), where \(A(1) = 1 - e_{1} + e_{2} = (1 - p_{1})(1 - p_{2})\) and \(D(1) = 1 - 2 \rho \cos\theta + \rho^{2}\).
% Since \(2 (e_{1} - e_{2} - 1) = -2 A(1)\), this is \(u = \left[D(1) - A(1)\right] / \left[2 A(1)\right] = (1 - \kappa) / (2 \kappa)\) with \(\kappa = A(1) / D(1)\), which is \cref{eq:kappa_osc} and which unity DC gain also makes \(1 / c_{0}\).
% Back-substituting \(u\) gives \(v\).
\begin{equation}\label{eq:uv_from_poles}
  u = \frac{1 - \kappa}{2 \kappa},
  \qquad
  v = \frac{2 u \left(p_{1} p_{2} - \rho \cos\theta\right) + p_{1} p_{2} - \rho^{2}}
           {2 \rho \sin\theta},
\end{equation}
where
\begin{equation}\label{eq:kappa_osc}
  \kappa = \frac{(1 - p_{1})(1 - p_{2})}{(1 - q_{1})(1 - q_{2})}
\end{equation}
is the DC-gain constant of \cref{eq:sos}, with denominator \((1 - q_{1})(1 - q_{2}) = 1 - 2 \rho \cos\theta + \rho^{2} \eqqcolon R^{2}\).
The in-phase component depends on the pole and zero locations only through \(\kappa\), so \(\delta u = -\delta \kappa / (2 \kappa^{2})\), and the logarithmic derivative of \cref{eq:kappa_osc} is
\begin{equation}\label{eq:dkappa_osc}
  \frac{\delta \kappa}{\kappa}
    = \sum_{i=1}^{2} \left(\frac{\delta q_{i}}{1 - q_{i}}
                           - \frac{\delta p_{i}}{1 - p_{i}}\right).
\end{equation}
% For a small residue, differentiating \cref{eq:uv_from_poles} and dropping the terms in \(u\) and \(v\) leaves \(\partial v / \partial u = (\rho - \cos\theta)/\sin\theta\), \(\partial v / \partial (p_{1} p_{2}) = 1/(2 \rho \sin\theta)\), and \(\partial v / \partial \rho = -1/\sin\theta\).
% \(\delta (p_{1} p_{2}) = 2 \rho_{p} \delta \rho_{p}\), and the same limit sends \(p_{i} \to q_{i}\) and so \(\rho_{p} / \rho \to 1\), leaving \(\delta \rho_{p} / \sin\theta\).
% Together, \(\delta v = (\rho - \cos\theta)\, \delta u / \sin\theta + (\delta \rho_{p} - \delta \rho) / \sin\theta\), so the quadrature component tracks the difference between the pole and zero radii.
% The bound \(V\) below is taken from the exact partials instead and reduces to this as \(\alpha_{r} \to 0\).
% \(\rho \to 1\) would make \(\partial v / \partial u\) into \(\tan(\theta/2)\), but nothing here holds \(1 - \rho\) below \(1 - \cos\theta\): at \(\rho \sin\theta = 1 - \rho^{2}\) the coefficient is \(-1/2\) while \(\tan(\theta/2)\) is \(+\theta/2\).
% \(\alpha_{r} = \sqrt{u^{2} + v^{2}}\) and \(\phi = \arctan(v, u)\), whose differentials are \cref{eq:osc_amp_phase}.
Finally,
\begin{equation}\label{eq:osc_amp_phase}
  \delta \alpha_{r} = \frac{u\, \delta u + v\, \delta v}{\alpha_{r}},
  \qquad
  \delta \phi = \frac{u\, \delta v - v\, \delta u}{\alpha_{r}^{2}}.
\end{equation}
With \(\abs{u}, \abs{v} \leq \alpha_{r}\), the triangle inequality applied to \cref{eq:dkappa_osc} and to the total differential of \(v\) in \cref{eq:uv_from_poles} gives \(\abs{\delta u} \leq U\) and \(\abs{\delta v} \leq V\), where
\begin{equation}\label{eq:osc_uv_bound}
  \begin{split}
    U &\coloneq \frac{\rho \mathcal{B}_{z}
                      + \rho_{p} \mathcal{B}_{p} / \sqrt{\kappa}}
                     {\kappa R},\\
    V &\coloneq \Gamma U
                + \frac{\rho_{p}^{2} \mathcal{B}_{p}}{\kappa \rho \sin\theta}
                + \Lambda \mathcal{B}_{z},
  \end{split}
\end{equation}
and
\begin{equation}\label{eq:osc_gamma_lambda}
  \begin{split}
    \Gamma &\coloneq \kappa \left(\frac{\abs{\rho - \cos\theta}}{\sin\theta}
                                  + 2 \alpha_{r}\right),\\
    \Lambda &\coloneq \frac{2 \alpha_{r} \abs{\cos\theta} + \rho}{\sin\theta}
                      + 2 \alpha_{r}.
  \end{split}
\end{equation}
\Cref{eq:osc_amp_phase} then gives
% \(\delta \left(\alpha_{r} e^{j \phi}\right) = e^{j \phi} \left(\delta \alpha_{r} + j \alpha_{r} \delta \phi\right)\), so the two terms are its radial and tangential components and their squares sum to \(\abs{\delta u + j \delta v}^{2}\).
% That is at most \(U^{2} + V^{2}\), and \(U\) and \(V\) are not attained together in any case, so \(\mathcal{E}\) remains conservative.
% Zeroing either term recovers \(\abs{\delta \alpha_{r}} \leq \mathcal{E}\) and \(\abs{\delta \phi} \leq \mathcal{E} / \alpha_{r}\), each attainable alone but not together, so treating them as independent overstates the residue displacement by \(\sqrt{2}\).
\begin{equation}\label{eq:osc_amp_phase_bound}
  \left(\delta \alpha_{r}\right)^{2} + \left(\alpha_{r} \delta \phi\right)^{2}
    \leq \mathcal{E}^{2},
  \qquad
  \mathcal{E} \coloneq \sqrt{U^{2} + V^{2}},
\end{equation}
where the two terms are the radial and tangential components of the residue displacement \(\delta u + j \delta v\).
% The \(\alpha_{r} \to 0\) limit reduces \(\mathcal{E}\) to \(\left(1 + 2/\kappa\right) \left(\mathcal{B}_{p} + \mathcal{B}_{z}\right) / \sin\theta\), but the reduction holds only in that limit, and reducing \(\mathcal{B}_{z}\) within it moves the notch of \(\Phi_{J}(q_{1})\) from \((\theta + \theta_{p}) / 2 = m \pi / J\) to \(\theta = m \pi / J\), understating \(\mathcal{E}\) between the two, so it is not used.
Two factors amplify \(\mathcal{E}\): \(1/\sin\theta\), which grows for a resonance near DC or Nyquist; and the peaks of \(\mathcal{B}_{p}\) at \(\theta_{p} = m \pi / J\) and of \(\mathcal{B}_{z}\) near \((\theta + \theta_{p}) / 2 = m \pi / J\), introduced by the transformation.

% TODO resurrect if we add bounce compensation support
% \subsection{Bounce Compensation}
% The parameters are \(\alpha_{e}\) and \(D\); \(D\) is an integer delay and is exact.
% TODO relate \(\delta \alpha_{e}\) to the truncated FIR taps \((-\alpha_{e})^{k}\) of \cref{eq:bounce}; the error compounds as \(k \alpha_{e}^{k-1} \delta \alpha_{e}\).
% TODO note that the time resolution of the delay is limited by the sample rate?

\section{Pole-Zero Cancellation}\label{sec:cancellation}

In the case of perfect precision, \(\Phi_{J}(z)\) contributes \(N_{a}(J-1)\) identical roots to \(A'(z)\) and \(B'(z)\), which cancel~\cite[Sec.~IV-B]{parhi1989pipeline}, causing the realized and ideal transfer functions to coincide: \(\hat{B}'(z) / \hat{A}'(z) = B(z) / A(z)\).
Coefficient quantization displaces those roots so that they no longer cancel, leaving \(N_{a}(J-1)\) pole-zero pairs in the realized transfer function~\cite[Sec.~II-B]{parhi1991finite}.
We quantify this effect by the fractional error, \(E(z)\), in the transfer function referred to the quantized poles:
\begin{equation}\label{eq:residual_definition}
  E(z) \coloneq \frac{\hat{B}'(z) / \hat{A}'(z)}{B(z) / \hat{A}(z)} - 1.
\end{equation}
That is, after quantizing \(A'(z)\), \(E(z)\) quantifies the additional error in the transfer function from quantizing \(B'(z)\).
% A step settles to the value its transfer function takes at \(\omega = 0\), which is \(E(1)\) for \(E\); the peak error elsewhere on the step is a partial sum of the impulse response of \(E\), which \(\lVert E \rVert_{\infty}\) does not bound.
% The \(n\)th partial sum of a function bounded by 1 on the unit disk is at most the Landau constant, which grows as \(\ln(n) / \pi\), so the peak step error over \(n\) samples is at most that constant times \(\lVert E \rVert_{\infty}\), a factor that diverges with the horizon.
% \(G \hat{B}' / \hat{A}' = G (1 + E) B / \hat{A} = (1 + E) A / \hat{A}\), since \(G B / A = 1\).
Since the correction imperfectly inverts the line, the cascade of the line distortion plus correction has response \((1 + E) A / \hat{A}\), where \(A / \hat{A}\) carries the pole displacement of \cref{sec:pole_sensitivity} and \(E\) is the fractional error that feedforward quantization adds at each frequency.
Referring to the quantized poles is appropriate because \cref{sec:pole_sensitivity} already accounts for their displacement.
Referring to them is also possible because \(\hat{A}'(z)\) contains \(z\) only in powers of \(z^{-J}\), so its \(N_{a} J\) roots form \(N_{a}\) sets of \(J\) equally spaced points for any \(\hat{a}_{k}'\).
% \(\left(1 - \hat{p}_{i} z^{-1}\right) \sum_{k=0}^{J-1} \hat{p}_{i}^{k} z^{-k} = 1 - \hat{p}_{i}^{J} z^{-J}\), the geometric sum, so \(\hat{A}\) divides \(\hat{A}'\) factor by factor and the quotient is \(\hat{\Phi}_{J}\).
Writing each set as the \(J\) rotations of a single \(\hat{p}_{i}\) gives \(\hat{A}'(z) = \prod_{i} \left(1 - \hat{p}_{i}^{J} z^{-J}\right) = \hat{A}(z) \hat{\Phi}_{J}(z)\), with
\begin{equation}\label{eq:ahat_phihat}
  \hat{A}(z) = \prod_{i=1}^{N_{a}} \left(1 - \hat{p}_{i} z^{-1}\right),
  \qquad
  \hat{\Phi}_{J}(z) = \prod_{i=1}^{N_{a}} \sum_{k=0}^{J-1} \hat{p}_{i}^{k} z^{-k}.
\end{equation}
By contrast, \(\hat{B}'(z) = \sum_{k=0}^{N_{b}'-1} \hat{b}_{k}' z^{-k}\) carries no such constraint and its roots are not tied to the \(\hat{p}_{i}\).
Perfect cancellation occurs when
\begin{equation}\label{eq:bprime}
  B'(z) = B(z) \hat{\Phi}_{J}(z).
\end{equation}
We therefore form \(B'(z)\) as this product and quantize its coefficients, rather than those of \(B(z) \Phi_{J}(z)\).
This places the added zeros directly on the added poles before quantization displaces them, removing the \(\Phi_{J} / \hat{\Phi}_{J}\) mismatch the latter would leave and carrying \(\hat{\Phi}_{J}\) into \cref{eq:zero_sensitivity}, a substitution of no consequence to first order.
% \(\hat{B}' = B' + \sum_{k} \delta b_{k}' z^{-k}\) and \(\hat{A}' = \hat{A} \hat{\Phi}_{J}\), so \(\hat{B}'/\hat{A}' = B/\hat{A} + \sum_{k} \delta b_{k}' z^{-k} / (\hat{A} \hat{\Phi}_{J})\).
% Dividing by \(B/\hat{A}\) cancels \(\hat{A}\) and leaves \(B \hat{\Phi}_{J} = B'\) in the denominator.
\Cref{eq:residual_definition} simplifies to
\begin{equation}\label{eq:residual}
  E(z) = \frac{\sum_{k=0}^{N_{b}'-1} \delta b_{k}' z^{-k}}{B'(z)}.
\end{equation}
Defining \(\lVert X \rVert_{\infty} \coloneq \max_{\omega} \abs{X(e^{j \omega})}\) for the worst case of a transfer function \(X\) over frequency, \(\lVert E \rVert_{\infty}\) is then bounded by
\begin{equation}\label{eq:residual_bound}
  \lVert E \rVert_{\infty}
    \leq \frac{N_{b}' \Delta b' / 2}{\min_{\omega} \abs{B'(e^{j \omega})}}.
\end{equation}
On the unit circle \(\abs{B'}\) is \(\abs{b'_{0}}\) times the product of the distances from \(e^{j \omega}\) to its roots, the zeros of \(B\) together with the added zeros at radius \(\abs{\hat{p}_{i}}\), so whichever root approaches the unit circle most closely dominates the minimum.
The subsections below evaluate that minimum at the design poles rather than the \(\hat{p}_{i}\), which shifts it by first order in \(\delta p_{i}\).

\subsection{High-Pass Droop}

% \(\abs{E(e^{j \omega})} = \abs{\delta b_{1}'} / \abs{1 - \rho e^{-j \omega}}\), whose denominator \(\sqrt{1 - 2 \rho \cos\omega + \rho^{2}}\) is least at \(\omega = 0\) because \(\rho > 0\).
\Cref{eq:high_pass_restrict_perturbation,eq:bprime} simplify \cref{eq:residual}, whose magnitude is greatest at \(\omega = 0\), so round to nearest gives
\begin{equation}\label{eq:integrator_residual}
  E(z) = \frac{\delta b_{1}' z^{-1}}{1 - \rho z^{-1}},
  \qquad
  \lVert E \rVert_{\infty} \leq \frac{\Delta b'}{2 \left(1 - \rho\right)}.
\end{equation}

\subsection{Exponential Tail}

% \(\hat{\Phi}_{J} = \left(1 - \hat{p}_{1}^{J} z^{-J}\right) / \left(1 - \hat{p}_{1} z^{-1}\right)\), so \cref{eq:tail_pre,eq:bprime} give \(B'(z) = H(z) \left(1 - p_{1}^{J} z^{-J}\right)\) with \(\hat{p}_{1}\) in place of \(p_{1}\), which the substitution above reads at \(p_{1}\).
% \(\abs{H}^{2}\) is monotone in \(\cos\omega\), its derivative there carrying the sign of \((p_{1} - q_{1})(1 - p_{1} q_{1})\), so its extrema are \(\abs{H(1)} = 1\), which \(1 - p_{1} = \kappa (1 - q_{1})\) fixes, and \(\abs{H(-1)} = \kappa (1 + q_{1}) / (1 + p_{1}) = (1 + \rho) / (1 + 2 \alpha + \rho)\).
% The latter is below 1 exactly when \(\alpha > 0\), so the smaller of the two is \((1 + \rho) / \left[1 + \rho + 2 \max(\alpha, 0)\right]\), whose denominator is positive without appeal to \cref{eq:exp_tail_alpha_restrict}.
% The second factor is minimized at the added zeros, where it is \(1 - \abs{p_{1}}^{J}\); \(\omega = 0\) is always one of them and \(\omega = \pi\) is whenever \(J\) is even, so the two minima then coincide and the product is the exact minimum.
% \(J T_{s} \ll (1 + \alpha) \tau\) is \(J (1 - p_{1}) \ll 1\), under which \(1 - \abs{p_{1}}^{J} \approx J (1 - \rho) / (1 + \alpha) \approx J T_{s} / \left[(1 + \alpha) \tau\right]\) and \(\rho \to 1\).
% Without that limit the reduction overstates \(1 - \abs{p_{1}}^{J}\), by \(J\) at \(p_{1} = 0\) and by more as \(p_{1} \to -1\), both of which \cref{eq:exp_tail_alpha_restrict} permits.
\Cref{eq:tail_pre,eq:bprime} factor \(B'\) into \(H\), whose magnitude is minimized at \(\omega = 0\) or \(\pi\), and \(1 - p_{1}^{J} z^{-J}\), minimized at the added zeros.
\(\min_{\omega} \abs{B'}\) is at least the product of their minima, so \cref{eq:residual_bound} gives
\begin{equation}\label{eq:tail_residual}
  \begin{split}
    \lVert E \rVert_{\infty}
      &\leq \frac{\left(J+1\right) \left[1 + \rho + 2 \max(\alpha, 0)\right] \Delta b'}
                 {2 \left(1 + \rho\right)
                  \left[1 - \abs*{\frac{\rho + \alpha}{1 + \alpha}}^{J}\right]}\\
      &\approx \frac{\left(J+1\right) \left(1 + \alpha\right)
                     \left[1 + \max(\alpha, 0)\right] \tau \Delta b'}
                    {2 J T_{s}},
  \end{split}
\end{equation}
where the simplification holds for \(J T_{s} \ll (1 + \alpha) \tau\).

\subsection{Damped Oscillation}

Here \cref{eq:sos,eq:bprime} factor \(B'\) into \(H\) and \(\hat{A}'\), both rational in the fitted poles and zeros, so \cref{eq:residual_bound} applies with \(\min_{\omega} \abs{B'}\) evaluated directly and \(N_{b}' = 2J + 1\).

\section{Q Format Derivations}\label{sec:q_format_derivations}

This appendix gives the derivations behind the requirements and results stated in \cref{sec:q_format}.

\subsection{Exponential Tail}

\Cref{eq:exp_stability,eq:tail_alpha_bound} constrain \(\Delta a'\), and \cref{eq:tail_alpha_bound,eq:tail_residual} constrain \(\Delta b'\).
The \(\tau\) and \(\alpha\) errors of \cref{sec:q_format} both follow from \(\delta q_{1}\), and for the same \(\delta q_{1}\) the peak step response error from \(\delta \tau\) is \(\abs{\alpha} / e\) times that from \(\delta \alpha\), so the \(\alpha\) requirement also holds the \(\tau\) error below \(t_{\mathrm{step}}\).
\Cref{eq:tail_bits} follows from \cref{eq:exp_pole_bound,eq:tail_zero_bound} in \cref{eq:tail_alpha_bound}, with \(\lambda\) apportioning \((1 + \alpha) t_{\mathrm{step}}\) between the \(\Delta a'\) and \(\Delta b'\) terms, and from \cref{eq:tail_residual}, under \(J T_{s} \ll (1 + \alpha) \tau\).
% Exactly, without the reduction, and with \(p_{1} = (\rho + \alpha) / (1 + \alpha)\) from \cref{eq:tail_zeros_poles}:
% \begin{equation}\label{eq:tail_bits_exact}
%   \begin{split}
%     \mathcal{S}_{+} &\coloneq \sum_{k=0}^{J} \rho^{-k},
%     \qquad
%     \mathcal{S}_{-} \coloneq \sum_{k=0}^{J-1} \left(\frac{p_{1}}{\rho}\right)^{k},\\
%     \mathcal{Z} &\coloneq \frac{\left(1 - \lambda\right) \left(1 - \rho\right)
%                                 t_{\mathrm{step}}}{\rho},\\
%     F_{a} &= \log_{2} \frac{\left(1 + \alpha\right)^{J}}
%                            {2 J \lambda \abs{\rho + \alpha}^{J-1}
%                             \left(1 - \rho\right) t_{\mathrm{step}}},\\
%     F_{b} &= \max \left\{
%       \log_{2} \frac{\left(1 + \alpha\right) \mathcal{S}_{+}}
%                     {2 \mathcal{Z} \abs{\mathcal{S}_{-}}},\;
%       \log_{2} \frac{\left(J + 1\right)
%                      \left[1 + \rho + 2 \max(\alpha, 0)\right]}
%                     {2 \left(1 + \rho\right)
%                      \left(1 - \abs{p_{1}}^{J}\right) t_{\mathrm{step}}}\right\}.
%   \end{split}
% \end{equation}
In comparison with the exact result, \cref{eq:tail_bits} understates \(F_{a}\) by \(\left[(J-1)/(1 + \alpha) + 1/2\right] T_{s} / (\tau \ln 2)\) bits and \(F_{b}\) by roughly half that.
% The penalty goes as \(1 / \left[(1 + \alpha) \tau\right]\), so it is worst at the most negative \(\alpha\) and the shortest \(\tau\); there the exact form agrees with the estimate to within a few hundredths of a bit.
At \(J = 8\) and \(T_{s} = \SI{1}{ns}\), this stays under 0.6 bits for \(F_{a}\) and 0.3 for \(F_{b}\) over \(\abs{\alpha} \leq 0.4\) and \(\tau \geq \SI{30}{ns}\).

\subsection{Damped Oscillation}

\Cref{eq:osc_stability_bound,eq:osc_amp_phase_bound} constrain \(\Delta a'\), and \cref{eq:osc_zero_bound,eq:osc_amp_phase_bound,eq:residual_bound} constrain \(\Delta b'\).
% Oscillation: differentiating \(g(t) = 1 + 2 \alpha_{r} e^{-t/\tau} \cos\psi\) gives \(\delta g = 2 \alpha_{r} (t / \tau) e^{-t/\tau} \left[\cos\psi\, \delta \tau / \tau + \sin\psi\, 2 Q\, \delta T_{\mathrm{osc}} / T_{\mathrm{osc}}\right]\), since \(2 \pi t / T_{\mathrm{osc}} = 2 Q t / \tau\); \(\mathcal{F} = -\delta G\, H\) with \(H \to 1\) makes the step error \(-\delta g\); the neglected factor is of relative size \(\alpha_{r} Q\) near resonance, which reaches about 7 over the range of \cref{tab:fixed_point}, so this is an approximation rather than an expansion.
% \((t / \tau) e^{-t/\tau}\) is largest at \(t = \tau\), where it is \(1/e\); the residue enters as \(\delta (2 \alpha_{r} e^{j \phi}) e^{-t/\tau} \cos\psi\), largest at \(t = 0\).
% The leading-order peaks are estimates, not bounds: over the damped oscillation range of \cref{tab:fixed_point}, the exact first-order step response of \(G \hat{H}\) exceeds the \(\tau\) and \(T_{\mathrm{osc}}\) expressions by up to a small factor, and the residue expression is accurate to a smaller one.
Approximating \(H\) by unity in \(\mathcal{F}\), the step response error is the error in the modeled step response of \cref{sec:damped_oscillation}, which with \(\psi \coloneq 2 \pi t / T_{\mathrm{osc}} + \phi\) is
\begin{equation}\label{eq:osc_step_error}
  -2 \alpha_{r} \frac{t}{\tau} e^{-t/\tau}
    \left[\cos\psi\, \frac{\delta \tau}{\tau}
          + \sin\psi\, 2 Q\, \frac{\delta T_{\mathrm{osc}}}{T_{\mathrm{osc}}}\right],
\end{equation}
so \(\delta \tau / \tau\) and \(\delta T_{\mathrm{osc}} / T_{\mathrm{osc}}\) produce peaks of \(2 \abs{\alpha_{r}} \abs{\delta \tau / \tau} / e\) and \(2 \abs{\alpha_{r}} (2 Q / e) \abs{\delta T_{\mathrm{osc}} / T_{\mathrm{osc}}}\) at \(t = \tau\), and the residue displacement \(\delta u + j \delta v\) of \cref{eq:osc_amp_phase_bound} produces \(2 \abs{\delta u + j \delta v} \leq 2 \mathcal{E}\) at \(t = 0\).
% \(\theta = 2 \pi T_{s} / T_{\mathrm{osc}}\) gives \(\delta \theta / \theta = -\delta T_{\mathrm{osc}} / T_{\mathrm{osc}}\) to first order, so \(2 Q\, \delta T_{\mathrm{osc}} / T_{\mathrm{osc}} = -(2 \pi \tau / T_{\mathrm{osc}}) \left[T_{\mathrm{osc}} / (2 \pi T_{s})\right] \delta \theta = -(\tau / T_{s}) \delta \theta\), and \cref{eq:tau_eps} gives \(\delta \tau / \tau \to (\tau / T_{s}) \delta \rho / \rho\).
% Cauchy--Schwarz on the bracket of \cref{eq:osc_step_error}: \(\abs{a \cos\psi + b \sin\psi} \leq \sqrt{a^{2} + b^{2}} = (\tau / T_{s}) \abs{\delta \rho / \rho + j \delta \theta} = (\tau / T_{s}) \abs{\delta q_{1} / q_{1}}\).
Separating \(\delta q_{1} / q_{1} = \delta \rho / \rho + j \delta \theta\) as in \cref{eq:osc_tau_tosc}, \(\delta \tau / \tau \approx (\tau / T_{s})\, \delta \rho / \rho\) and \(2 Q\, \delta T_{\mathrm{osc}} / T_{\mathrm{osc}} \approx -(\tau / T_{s})\, \delta \theta\), so the bracket of \cref{eq:osc_step_error} is at most \((\tau / T_{s}) \mathcal{B}_{z}\) and the combined peak step response error from \(\tau\) and \(T_{\mathrm{osc}}\) is \((2 \abs{\alpha_{r}} / e) (\tau / T_{s}) \mathcal{B}_{z}\), the same as either alone.
Requiring each to be at most \(t_{\mathrm{step}}\) gives
\begin{equation}\label{eq:osc_step_tolerances}
  \mathcal{B}_{z} \leq \frac{e\, t_{\mathrm{step}} T_{s}}{2 \abs{\alpha_{r}} \tau},
  \qquad
  \mathcal{E} \leq \frac{t_{\mathrm{step}}}{2}.
\end{equation}
We also require \(\lVert E \rVert_{\infty} \leq t_{\mathrm{step}}\) in \cref{eq:residual_bound}.

Three factors grow without bound: \(1/\sin\theta\) in \cref{eq:osc_gamma_lambda} for a resonance near DC or Nyquist, \(1/\abs{\sin (J \theta_{p})}\) in \cref{eq:osc_pole_bound} at \(\theta_{p} = m \pi / J\), and \(1 / \abs{\Phi_{J}(q_{1})}\) in \cref{eq:osc_zero_bound} at \((\theta + \theta_{p}) / 2 = m \pi / J\).
% Both read \(J \omega = m \pi\), \(J\) samples spanning \(m\) half periods at \(\omega\), and differ only in the \(\omega\) taken.
The last two arise from the feedback delay spanning a multiple of a half period, of the poles in one case and of the mean of the poles and zeros in the other, so they coincide only where \(\theta_{p} = \theta\).
% Within \(W\) of the degeneracy the square root branch of \cref{eq:osc_pole_bound} governs, and inverting it doubles the bits a given accuracy costs.
\Cref{eq:osc_stability_bound} collapses at those same \(\theta_{p}\), where its first term is \((1 - \sigma)^{2}\), which vanishes as \(\sigma\) approaches unity for a long-lived resonance.
Each of these drives the required coefficient width up without bound, so no finite width covers the whole parameter space.

% \(D - \hat{A} = 2 \Delta_{1} z^{-1} - \Delta_{2} z^{-2}\) with \(D\) the zero polynomial of \cref{eq:sos}, so \(B' = \kappa D \Phi_{J} = \kappa \left[A' + \left(2 \Delta_{1} z^{-1} - \Delta_{2} z^{-2}\right) \Phi_{J}\right]\).
% \(A'\) carries only \(1\), \(-a_{1}'\), and \(-a_{2}'\), which the paragraph above bounds by 2.
% Over the range below, \cref{eq:osc_ib} stays under 3 against the 4 that \(I_{b} = 3\) carries, and the realized \(\max_{k} \abs{b_{k}'}\) stays under 2.5; \(I_{b} = 2\) would hold only \(\abs{b_{k}'} < 2\), which \cref{eq:osc_ib} reaches at zero residue.
Subtracting \(\hat{A}\) from the zero polynomial of \cref{eq:sos} splits \(B'\) into \(\hat{A} \Phi_{J} = A'\) and a remainder driven by the pole-zero mismatch, giving
\begin{equation}\label{eq:osc_ib}
  \max_{k} \abs{b_{k}'}
    \leq \kappa \left[2
      + \Psi \left(2 \abs{\Delta_{1}} + \abs{\Delta_{2}}\right)\right],
\end{equation}
where \(\Delta_{1} \coloneq \operatorname{Re}(p_{1} - q_{1})\) and \(\Delta_{2} \coloneq \rho_{p}^{2} - \rho^{2}\) measure that mismatch and \(\Psi \coloneq \min\left(J, 1/\abs{\sin\theta_{p}}\right)\) bounds the coefficients of \(\Phi_{J}\), since the solution of \cref{eq:osc_h_recursion} is \(h_{m} = \rho_{p}^{m} \sin\left[(m+1) \theta_{p}\right] / \sin\theta_{p}\).
The leading 2 is \(\max_{k} \abs{a_{k}'}\).

% \cref{eq:osc_pole_bound,eq:osc_zero_bound} need only a conjugate pair inside the unit circle, so they carry to any residue.
A section outside the range of \cref{sec:q_osc} is correctable whenever it satisfies \cref{eq:osc_stability_bound}, \(\max_{k} \abs{b_{k}'} < 4\) for Q3.24, \(\min_{\omega} \abs{B'} \geq 2^{-11}\) for Q1.26, \cref{eq:osc_step_tolerances}, and \(\lVert E \rVert_{\infty} \leq t_{\mathrm{step}}\) by \cref{eq:residual_bound}.

\fi % end of the long-version appendix

\section*{Acknowledgment}
This material was funded by the U.S. Department of Energy, Office of Science, Office of Advanced Scientific Computing Research Quantum Testbed Program under contract DE-AC02-05CH11231.

Anthropic's Claude Opus 5 and Claude Fable 5.1 assisted with derivations in \cref{sec:fixed_point}\longonly{ and the appendices}, and with simulating the peak step response error for each distortion model over the specified parameter ranges.

\bibliographystyle{IEEEtran}
\bibliography{biblio}

\end{document}